\documentclass[journal,twoside,web]{ieeecolor}
\usepackage{tmi}
\usepackage{cite}
\usepackage{amsmath,amssymb,amsfonts}
\usepackage{algorithmic}
\usepackage{graphicx}
\usepackage{textcomp}
\usepackage{multirow}
\usepackage{booktabs}
\usepackage{color}
\usepackage{bm}
\usepackage{hyperref}
\usepackage{caption}
\usepackage{amsmath}
\usepackage[linesnumbered,ruled]{algorithm2e}
\def\BibTeX{{\rm B\kern-.05em{\sc i\kern-.025em b}\kern-.08em
    T\kern-.1667em\lower.7ex\hbox{E}\kern-.125emX}}
\begin{document}
\title{Continuous 3D Myocardial Motion Tracking via Echocardiography}

\author{Chengkang Shen, Hao Zhu, You Zhou, Yu Liu, Si Yi, Lili Dong, Weipeng Zhao, David J. Brady, Xun Cao, Zhan Ma, and Yi Lin
\thanks{This work was supported by the NSFC (No. 62331005, 62071219, 62025108) and the Science and Technology Commission of Shanghai Municipality (STCSM) (No. 20Y11909800).}
\thanks{Chengkang Shen, Hao Zhu, Si Yi, Xun Cao and Zhan Ma are with the School of Electronic Science and Engineering, Nanjing University, Nanjing, 210023, China. }
\thanks{You Zhou is with the Medical School, Nanjing University, Nanjing, 210093, China.}
\thanks{Yu Liu, Lili Dong and Weipeng Zhao are with the Department of Echocardiography of Zhongshan Hospital, Fudan University, Shanghai, 200032, China.}
\thanks{David J. Brady is with the College of Optical Sciences, University of Arizona, Tucson, AZ 85721, USA.}
\thanks{Yi Lin is with the Department of Cardiovascular Surgery of Zhongshan Hospital, Fudan University, Shanghai, 200032, China.}
\thanks{Chengkang Shen, Hao Zhu and You Zhou contributed equally to this work.}
\thanks{Corresponding authors: Yi Lin, Zhan Ma, and Xun Cao. (e-mail: lyricist2012@163.com; mazhan@nju.edu.cn; caoxun@nju.edu.cn)}
\thanks{This work has been submitted to the IEEE for possible publication. Copyright may be transferred without notice, after which this version may no longer be accessible.}}
\maketitle

\begin{abstract}
Myocardial motion tracking stands as an essential clinical tool in the prevention and detection of cardiovascular diseases (CVDs), the foremost cause of death globally. However, current techniques suffer from incomplete and inaccurate motion estimation of the myocardium in both spatial and temporal dimensions, hindering the early identification of myocardial dysfunction. To address these challenges, this paper introduces the Neural Cardiac Motion Field (NeuralCMF). NeuralCMF leverages implicit neural representation (INR) to model the 3D structure and the comprehensive 6D forward/backward motion of the heart. This method surpasses pixel-wise limitations by offering the capability to continuously query the precise shape and motion of the myocardium at any specific point throughout the cardiac cycle, enhancing the detailed analysis of cardiac dynamics beyond traditional speckle tracking.
Notably, NeuralCMF operates without the need for paired datasets, and its optimization is self-supervised through the physics knowledge priors in both space and time dimensions, ensuring compatibility with both 2D and 3D echocardiogram video inputs. Experimental validations across three representative datasets support the robustness and innovative nature of the NeuralCMF, marking significant advantages over existing state-of-the-art methods in cardiac imaging and motion tracking. Code is available at: \href{https://njuvision.github.io/NeuralCMF/}{https://njuvision.github.io/NeuralCMF}.

\end{abstract}

\begin{figure*}[ht!]

\begin{center}
\includegraphics[width=\textwidth]{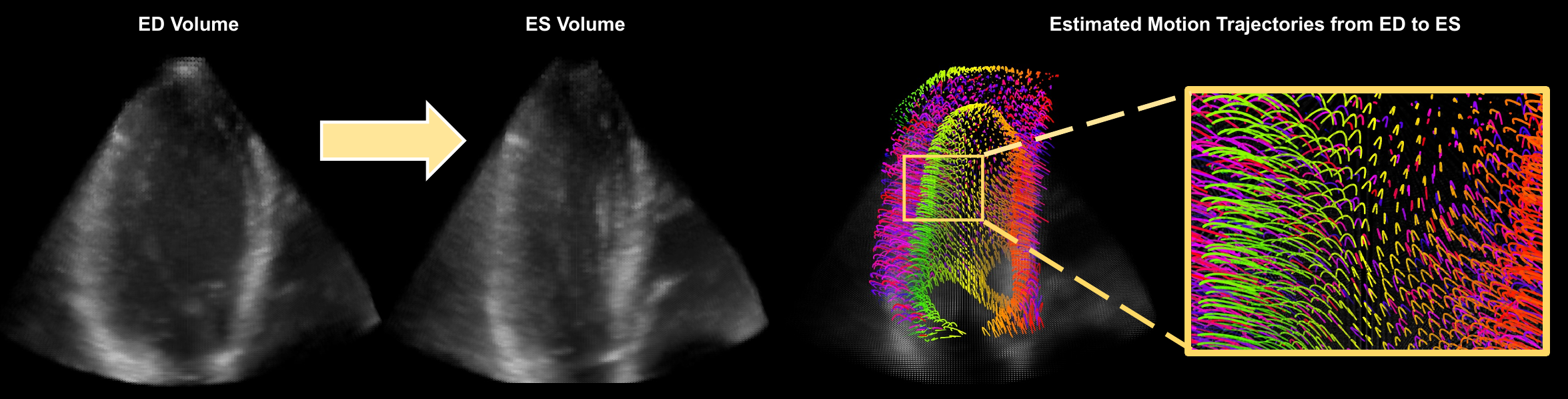}
\end{center}
\caption{We introduce a self-supervised method for estimating 3D motion trajectories throughout the cardiac cycle at any given point all at once. The figure illustrates the motion estimation of the 3D heart from end-diastole (ED) to end-systole (ES), representing the heart's contraction phase. The estimated motion trajectories are displayed on the right with each point's trajectory depicted in a different color for clarity. Though we have chosen to show only sparse trajectories of the left ventricular (LV) myocardium, our method has the ability to compute motion for all points across the entire 3D heart.}
\label{flow_results}
\end{figure*}

\section{Introduction}
\label{sec:introduction}
\IEEEPARstart{C}{ardiovascular} diseases (CVDs) stand as the primary global cause of death \cite{nowbar2019mortality}. While CVDs encompass a spectrum of pathological disorders, including coronary atherosclerosis, valvular insufficiency, and inflammatory myocarditis, each impairs the heart's motion and pumping function \cite{tsao2022heart}. Therefore, accurate tracking of myocardial motion is pivotal for detecting CVDs. For instance, motion tracking is critical for early intervention in cardiotoxicity by detecting subclinical cardiac dysfunction post-cancer treatment \cite{zhang2018abnormalities}, identifying regional contractility inconsistencies in ischemic diseases for coronary artery disease detection \cite{van2019adding}, and assessing risks and treatment success in valvular heart diseases \cite{alashi2018incremental, kim2018myocardial}. It also aids in the early detection of systolic and diastolic dysfunction in cardiomyopathies, forecasting cardiovascular events and patient outcomes \cite{morris2017potential, park2018global}, optimizing cardiac resynchronization therapy \cite{risum2012simple}, understanding hemodynamics in pulmonary hypertension and congenital heart diseases \cite{fine2013outcome, chowdhury2015speckle}, and managing atrial fibrillation \cite{leung2018left,motoki2014global}. Overall, cardiac motion tracking is indispensable across a wide range of heart diseases for its diagnostic, prognostic, and therapeutic guidance capabilities.
Currently, two primary modalities for myocardial motion tracking exist, namely echocardiography and cardiac magnetic resonance (CMR). Considering factors such as economics, prevalence, operational complexity, and temporal resolution, echocardiography is favored over CMR \cite{braga2019trends,liu2019deep,voigt20192}. Consequently, developing and implementing an accurate and robust 3D myocardial motion tracking method based on echocardiography is essential in modern cardiology practice.

While numerous methods have been proposed to track myocardial motion in echocardiogram video, several challenges persist in existing solutions. In speckle-based tracking approaches \cite{mondillo2011speckle, lubinski1999speckle, chen20053, voigt20192} which aim to identify consistent intensity patterns across successive frames by optimizing their similarities, capturing motion trajectories over extended temporal frames remains a formidable task due to the inherent limitation of estimating motions frame by frame. Surface-based tracking techniques \cite{papademetris2002estimation, parajuli2019flow, huang2014contour, parajuli2016integrated, shi2000point, lin2004generalized}, on the other hand, concentrate on capturing similar curvature or shape characteristics of the chamber's boundaries across sequential frames. However, the necessity for interpolating motion vectors for points within the boundary introduces inaccuracies, particularly in myocardium regions that are not fully infarcted. For instance, in cases of cardiac sarcoidosis \cite{blankstein2016evaluation, ichinose2008mri}, where motion across the entire myocardium lacks uniformity. More recently, deep learning-based tracking methods have achieved remarkable success in analyzing deformations between images \cite{ahn2020unsupervised,ta2020semi,ahn2023co}. However, they face limitations in predicting the complete cardiac cycle's motion and suffer from the lack of large-scale paired datasets. In summary, current myocardial motion tracking techniques with echocardiogram video input encounter three primary challenges: (1) \textit{accurate and complete estimation of myocardial motion throughout the entire cardiac cycle}, (2) \textit{tracking any points in the 3D heart without resorting to interpolation}, 
and (3) \textit{operating without the need for additional paired training datasets.}

In this manuscript, we present Neural Cardiac Motion Field (NeuralCMF), an innovative self-supervised technique designed to simultaneously model the 3D cardiac structure and capture the 6D forward/backward motions of the myocardium. In contrast to prior methods that primarily focus on comparing neighboring frames pairwise, NeuralCMF takes a unique approach by modeling the global motion of all pixels throughout the entire video as a continuous implicit neural function, which has a coordinate input and the corresponding shape/motion output. Consequently, this enables us to precisely query the motion trajectory for any pixel spanning the complete cardiac cycle, as demonstrated in Fig. \ref{flow_results}. Rather than depending on an extensive paired dataset for supervision, NeuralCMF employs a physics-informed self-supervision mechanism. This mechanism involves comparing the loss between the network output after physical imaging operators and the input videos during optimization (see Fig. \ref{pipeline}). As a result, NeuralCMF effectively overcomes the dataset dependency commonly found in current deep learning-based methods, eliminating the limitations posed by data bias and enhancing its applicability and generalizability across a broad spectrum of ultrasound devices and patient populations.
Furthermore, as the physical imaging operators can be modified according to 2D echocardiogram (2DE) or 3D echocardiogram (3DE) videos, NeuralCMF can be seamlessly integrated with both multi-view 2DE and 3DE video inputs. 
To the best of our knowledge, this universality for estimating 3D myocardial motion using echocardiogram videos from both 2D and 3D sources is a first in the community. The quantitative and qualitative experiments conducted on three representative datasets, spanning both 2DE and 3DE videos, show that the performance of our approach consistently outperforms the existing state-of-the-art methods \cite{balakrishnan2019voxelmorph, ahn2023co}. We believe the methodology presented here may also be applicable across more imaging modalities beyond ultrasonic imaging for analyzing the motion of organs other than the heart.

The contributions of this paper are summarized below:
\begin{itemize}
\item We present NeuralCMF, a novel self-supervised method
that simultaneously models the 3D structure of the heart
and the 6D forward/backward motion of the myocardium. This method employs a personalized training strategy, where the network is specifically trained using the heart video data of an individual. With this approach, NeuralCMF is capable of estimating the 3D motion of any specific point in the heart throughout the cardiac cycle.
\item The capability of NeuralCMF to estimate dense 3D motion from multi-view 2DE images represents both innovation and practicality, rendering it particularly valuable in diverse critical care environments such as emergency departments and preoperative clinics.
\item To evaluate the performance and validity of our method, we have established unique multi-view 2DE and 3DE video datasets for cardiac imaging comprising data from 127 volunteers, covering both healthy individuals and patients with coronary heart disease (CHD).
\item NeuralCMF provides state-of-the-art performance in quantitative and qualitative assessments tested on three separate datasets: the open-source STRAUS \cite{alessandrini2015pipeline}, along with our specially constructed multi-view 2DE and 3DE video datasets.
\end{itemize}

\section{Related Work}
\subsection{Cardiac Motion Tracking Methods}
Motion tracking is crucial for precise strain calculations, and its objective is to map the displacement between consecutive images in a sequence. This section reviews existing motion tracking algorithms.
\subsubsection{Intensity-based Methods}
Speckle-tracking methods identify consistent intensity patterns across several consecutive time frames by maximizing the similarity of these patterns \cite{lubinski1999speckle, chen20053}. These algorithms operate under the premise that there is only a small displacement motion between frames and that the intensity of a specific point within a moving object will remain unchanged. Nevertheless, the precision of speckle tracking tends to diminish when the temporal intervals between frames become substantial.

\subsubsection{Feature-based Methods}
Feature-based algorithms utilize image attributes such as surfaces, texture, or curvature, allowing consistent tracking over time \cite{papademetris2002estimation, huang2014contour, shi2000point}. These methods generate motion fields by matching the key points across frames based on features. The point matching algorithms include iterative closest point matching \cite{besl1992method} and robust point matching \cite{chui2003new}. Although efficient, these approaches are limited by the dependency on accurate segmentation and often produce sparse motion fields that need additional interpolation to generate motion in the desired locations.
\subsubsection{Deep Learning-based Methods}
State-of-the-art methods are driven by deep learning, gaining traction for their computational efficiency and pattern recognition capabilities. In the realm of supervised learning, a variety of methods have emerged: Roh\'e \textit{et al.} \cite{rohe2017svf} developed a method that leveraged a convolutional neural network (CNN) guided by the mesh segmentation of paired images to compute transformations between pairs of CMR images. Parajuli \textit{et al.} \cite{parajuli2019flow}  introduced the flow network tracker (FNT), employing a convolutional autoencoder to extract a point cloud from myocardium segmentation, followed by point tracking across time frames. Wu \textit{et al.} \cite{wu2018deep} utilized a deep boltzmann machine (DBM) to capture both local and global heart shape variations, enabling frame-by-frame heart motion tracking in CMR images.
However, these supervised methodologies predominantly rely on simulated datasets with known ground-truth deformations or expert-labeled myocardium segmentations, which are difficult to collect in real cases.

Limited ground truth motion labels have increased interest in unsupervised motion tracking. This has led to the evolution of methods employing diverse U-Net architectures \cite{ronneberger2015u} for cardiac motion tracking  \cite{ahn2020unsupervised, yu2020foal, yu2020motion}. However, these methods have mostly been limited to the analysis of 2D heart images or CMR images, leaving motion tracking on 3DE an enduring challenge. Recent works, such as Ta \textit{et al.} \cite{ta2020semi} with a proposed semi-supervised joint network and Ahn \textit{et al.} \cite{ahn2023co} with a co-attention spatial transformer network, both for pairs of 3D heart images motion estimation, have shown promising advancements. Yet, these approaches remain limited in concurrently predicting the entire cardiac cycle's motion, necessitating extensive training datasets and exhibiting susceptibility to data bias.

Deep learning-based methods have shown significant promise in cardiac motion tracking. Nonetheless, no methods can predict complete myocardial motion throughout the cardiac cycle without relying on simulated or expert-labeled datasets, a factor that is crucial for clinical use.

\subsection{Implicit Neural Representation}
The implicit neural representation (INR) \cite{sitzmann2020implicit}, also referred to as neural fields \cite{mildenhall2021nerf}, characterizes the function of a signal by mapping the coordinate to its attribute using a neural network. Different from previous deep-learning techniques \cite{he2016deep}, which are supervised by a large-scale paired dataset, the training of INR follows a self-supervised mechanism. This self-supervised mechanism can be embedded with various differentiable physics operators, making the INR naturally suitable for solving domain-specific inverse problems where large-scale paired datasets are unavailable. As a result, INR has been widely applied in various tasks, such as the neural rendering \cite{mildenhall2021nerf, zhu2023pyramid}, microscopy imaging \cite{liu2022recovery, zhu2022dnf}, material design \cite{chen2020physics} and partial differential equations solver \cite{karniadakis2021physics}. 

Recently, INR has gained prominence in medical imaging. 
Focusing on the realm of 2D/3D image reconstruction, the prevalent modalities such as computed tomography (CT), magnetic resonance imaging (MRI), and ultrasound have seen numerous advancements. For instance, in the context of CT image reconstruction, Shen \textit{et al.} \cite{shen2022nerp} introduced a method that used prior images and INR to reconstruct high-quality 2D CT images from sparse samples. Zha \textit{et al.} \cite{zha2022naf} proposed a method for sparse-view cone beam computed tomography (CBCT) reconstruction, using INR to model the attenuation field and optimize it by minimizing discrepancies between actual and synthesized projections. In the MRI reconstruction realm, Wu \textit{et al.}  \cite{wu2022arbitrary} used high and low-resolution MRI image pairs to represent the 3D brain surface, processed the low-resolution images with a super-resolution CNN, and combined the extracted features with high-resolution coordinates in an MLP decoder to determine 3D voxel intensity. For ultrasound reconstruction, Shen \textit{et al.} \cite{shen2023cardiacfield} proposed CardiacField to produce a high-fidelity 3D heart and automatically estimate heart function using only low-cost 2DE probes.

While numerous methods have employed INR to address challenges in medical imaging, none have explored its application for cardiac motion analysis in ultrasound modality, which is paramount in the clinical assessment of cardiac function.

\section{Methodology}
\begin{figure*}[ht!]
\begin{center}
\includegraphics[width=\textwidth]{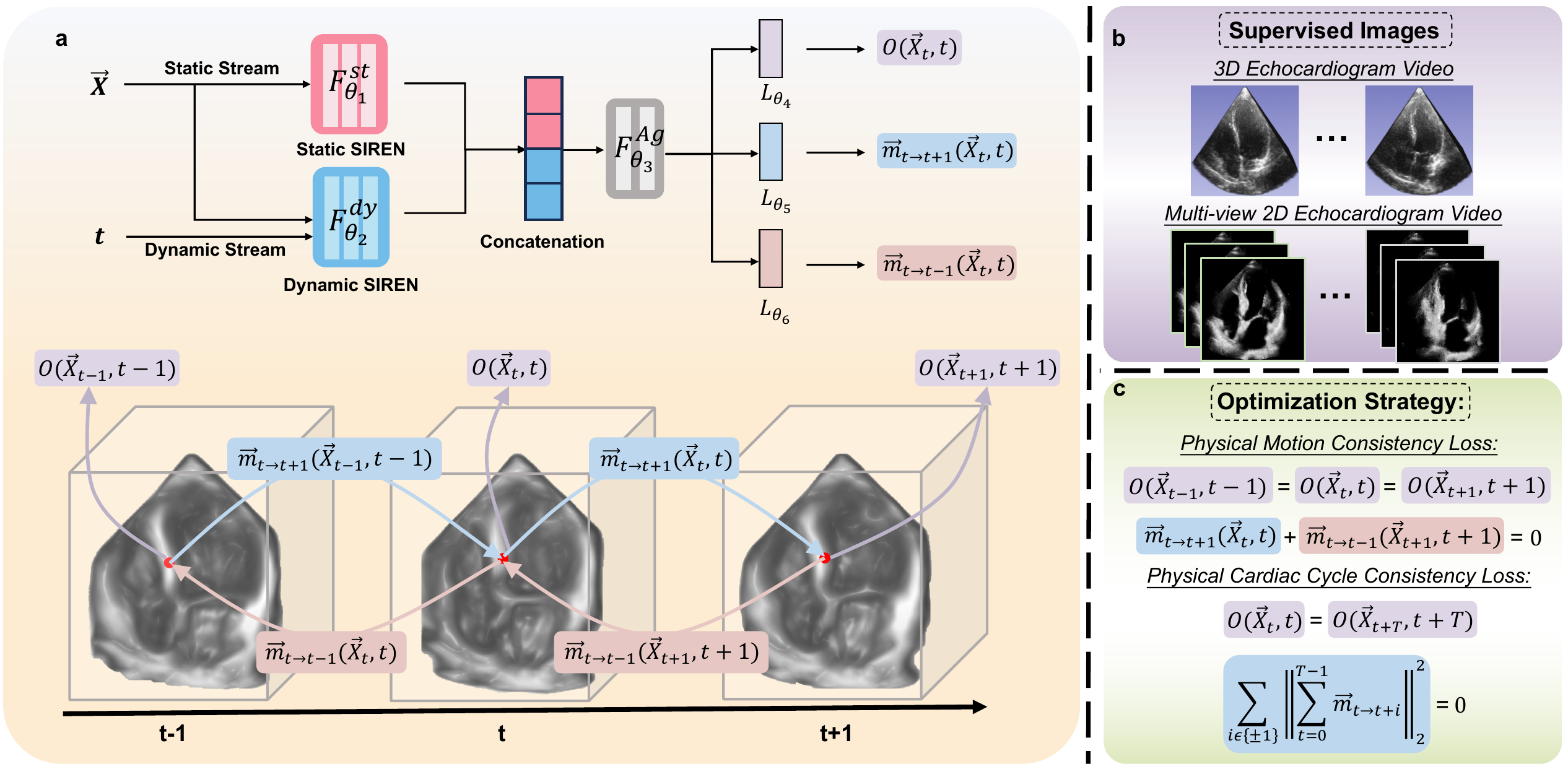}
\end{center}
\caption{The figure illustrates the workflow of our method. (a) Inputs consist of 4D space-time coordinates $(x,y,z,t)$. Corresponding outputs include the intensity value ${O}(x,y,z,t)$ at specified space-time locations, as well as forward and backward 3D motion vectors ${\vec{M}}=(\vec{m}_{t \rightarrow t+1},\vec{m}_{t \rightarrow t-1})$ that represent the movements of $(x,y,z)$ at successive time frames $t+1$ and  $t-1$, respectively. (b) Our approach can be applied to both 2DE and 3DE videos. (c) The methodology utilizes two main optimization strategies: physical motion consistency loss, ensuring that intensity remains constant within both the heart's tissues and blood-filled areas throughout the entire cardiac cycle, and physical cardiac cycle consistency loss, which aligns the process with the physiological fact that the heart muscle returns to its initial shape upon completing one full cardiac cycle.}
\label{pipeline}
\end{figure*}

\subsection{Overview}
Aligned with our aim to attain a comprehensive understanding of cardiac dynamics, we present a self-supervised methodology that utilizes INR to accurately estimate the 3D motion of the myocardium throughout the entire cardiac cycle. The method is applicable to both 2DE and 3DE videos. As depicted in Fig. \ref{pipeline},  the inputs of our method are the 4D space-time coordinates $(x,y,z,t)^{\top}$ in the heart. The corresponding outputs consist of the intensity value ${O}(x,y,z,t)$ at the specified space-time location, along with the forward and backward motion vectors ${\vec{M}}=\{\vec{m}_{t \rightarrow t+1},\vec{m}_{t \rightarrow t-1}\}$ of the location $(x,y,z,t)^{\top}$ at successive time frames $t+1$ and  $t-1$, respectively.

Our approach extends our previous work, CardiacField \cite{shen2023cardiacfield}, by integrating the temporal dimension into the network input, enabling the quantification of the motion of the myocardium throughout the entire cardiac cycle. In the following sections, we will begin by briefly reviewing the core of the CardiacField \cite{shen2023cardiacfield}. Subsequently, we will detail the NeuralCMF and its optimization strategy.

\subsection{Background}
Our previous study, CardiacField \cite{shen2023cardiacfield}, focuses on reconstructing the 3D heart from a multi-view 2DE video captured from different angles. To achieve this, it represents a static 3D heart as a continuous function using the INR, denoted by $F_{\theta}^{static}$. This function takes 3D positions $\vec{X}=(x,y,z)^{\top}$ as input and yields the corresponding volume intensity ${O}(\vec{X})$,
 \begin{equation} 
    {O}(\vec{X}) = F_{\theta}^{static}(\vec{X}),
\end{equation}
where the function $F_{\theta}^{static}$ is modeled using a multi-resolution hash-based MLP network \cite{muller2022instant, zhu2023disorder}. To supervise the training of the network parameters in CardiacField \cite{shen2023cardiacfield}, the physical imaging process of 2DE images is introduced to construct the physics-informed loss function. In detail, each 2DE image can be viewed as applying a virtual slicing operation to the 3D cardiac volume according to the angles of different images, \textit{i.e.},
 \begin{equation} 
    \mathbf{I}_s = \mathcal{S}_{2}^{3} \circ \mathcal{B}\circ {O}(\vec{X}),
    \label{eqn:slicing_imaging}
\end{equation}
where $\mathcal{S}_{M}^{N}$ is the slicing operator that reduces an $N$-dimensional function down to an $M$-dimensional one by zeroing out the last $N-M$ dimensions. $\mathcal{B}$ characterizes the angle of different images and is used to transform the coordinates from the global world coordinate system to the image plane system.
Subsequently, the 2DE images ${\{\mathbf{I}_{s}^{i}\}}_{i=1}^{k}$ derived from ${O}(\vec{X})$, are compared to ${\{\mathbf{I}_{g}^{i}\}}_{i=1}^{k}$ in the input 2DE images using a mean squared error loss.
\begin{equation} 
    \mathcal{L}_{CardiacField}=\frac{1}{K} \sum_{i}^{K}\left\lVert \text{vec}(\mathbf{I}_s^{i}) - \text{vec}(\mathbf{I}_g^{i}) \right\rVert^2_{2},
\label{eqn:loss_phy_loss}
\end{equation}
where $K$ refers to the number of input 2DE images, $\mathbf{I}_{s}^{i}$ and $\mathbf{I}_{g}^{i}$ denote the $i$-th synthesized image from the network and the corresponding ground truth input image.
The $\lVert \cdot \rVert_{2}$ denotes the $L_2$-norm of the vector or matrix. 
The $\text{vec}(\cdot)$ denotes the vectorization of a matrix.
Finally, the loss $\mathcal{L}_{CardiacField}$ is used to update the network parameters $\theta$ in $F_{\theta}^{static}$ and the angle parameters in $\mathcal{B}$ following the backward gradient flow. Once the training of $F_{\theta}^{static}$ converges, the 3D structure of the heart could be reconstructed by feeding the coordinate $\vec{X}$ to the $F_{\theta}^{static}$ one by one. As a result, CardiacField could be used to reconstruct the 3D heart from 2DE video without attaching any other position sensors.

\subsection{Neural Cardiac Motion Field}
In order to quantify the deformation of the myocardium, we extend the static 3D heart representation in CardiacField \cite{shen2023cardiacfield} to dynamic reconstruction and formulate the dense 3D motion field of the myocardium using the INR. Given a 3D position $\vec{X}$ at a specific time $t$, our model $F_{\theta}^{Motion}$ predicts not only the volume intensity ${O}(\vec{X}_{t},t)$ but also the forward and backward motion vectors. These vectors are expressed as ${\vec{M}(\vec{X}_{t},t)}=\{\vec{m}_{t \rightarrow t+1}(\vec{X}_{t},t),\vec{m}_{t \rightarrow t-1}(\vec{X}_{t},t)\}$. Specifically, the forward 3D offset vector  $\vec{m}_{t \rightarrow t+1}(\vec{X}_{t},t)$ is represented as $(x_f,y_f,z_f)^{{\top}}$, detailing the displacement from time frame $t$ to $t+1$. Similarly, the backward 3D offset vector $\vec{m}_{t \rightarrow t-1}(\vec{X}_{t},t)$ is denoted as $(x_b,y_b,z_b)^{{\top}}$, which encapsulates the displacement from time frame $t$ to $t-1$. These vectors provide a comprehensive description of the motion in 6D, accounting for spatial displacement in three dimensions over two time intervals.
Accordingly, the dynamic model is mathematically defined as follows,
\begin{equation}
    \{{O}(\vec{X}_{t},t), {\vec{M}(\vec{X}_{t},t)} \}= F_{\theta}^{Motion}(\vec{X}_{t},t).
    \label{eqn:fw}
\end{equation}

As depicted in Fig. \ref{pipeline}, the network $F_{\theta}^{Motion}$ consists of a static and dynamic stream. The static stream takes the 3D position vector $\vec{X}_{t}$ as its input, while the dynamic stream's input consists of the concatenation of the 3D position vector $\vec{X}_{t}$ and time $t$. The networks within the static $F_{\theta_{1}}^{st}$ and dynamic $F_{\theta_{2}}^{dy}$ streams are formulated using the SIREN network \cite{sitzmann2020implicit}. Each network has 4 layers, with 64 neurons in each hidden layer. The outputs of the two streams are static and dynamic feature vectors with length 32, respectively. The static feature vector captures the heart's shape in the first frame. The dynamic feature vectors are essential to capture the transient changes of the myocardium that occur over time. These feature vectors are then concatenated and fed into a feature aggregation network $F_{\theta_{3}}^{Ag}$, comprising a single layer with 64 neurons. The output $\mathbf{H}$ of the feature aggregation network is further fed into three distinct layers ($L_{\theta_{4}}$, $L_{\theta_{5}}$, and $L_{\theta_{6}}$) to obtain the volume intensity ${O}(\vec{X}_{t},t)$ and motion vectors $\vec{m}_{t \rightarrow t+1}(\vec{X}_{t},t)$ and $\vec{m}_{t \rightarrow t-1}(\vec{X}_{t},t)$,
\begin{equation}
\begin{aligned}
    \mathbf{H}=F_{\theta_{3}}^{Ag}(\text{Concat}&(F_{\theta_{1}}^{st}(\vec{X}_{t}),F_{\theta_{2}}^{dy}(\vec{X}_{t},t))) \\
    {O}(\vec{X}_{t},t) &= \sigma(L_{\theta_{4}}(\mathbf{H})) \\
    \vec{m}_{t \rightarrow t+1}(\vec{X}_{t},t) &= \tanh(L_{\theta_{5}}(\mathbf{H})) \\
    \vec{m}_{t \rightarrow t-1}(\vec{X}_{t},t) &= \tanh(L_{\theta_{6}}(\mathbf{H}))
\end{aligned},
\end{equation}
where $\theta_{i}$ denotes the parameters of the corresponding neural network. $\sigma$ and $\tanh$ refer to the Sigmoid and Tanh activation functions, respectively.

\subsection{Physics-informed Optimization}
To supervise the training of the network parameters, a physics-informed loss function is proposed,
\begin{equation} 
    \mathcal{L} = \underbrace{\alpha_1\mathcal{L}_{image}}_{Static\: Prior} + \underbrace{\mathcal{L}_{motion} + \mathcal{L}_{cycle}}_{Motion\: Prior} + {\alpha_3 \mathcal{L}_{reg}}
\end{equation}
where $\mathcal{L}_{image}$ and $\mathcal{L}_{motion}$, $\mathcal{L}_{cycle}$ respectively model the static and dynamic physical knowledge priors, $\mathcal{L}_{reg}$ regularizes the loss function. The terms  $\alpha_1$ and $\alpha_3$ represent the weight of the physical imaging loss and regularization loss.
Subsequent sections will provide a detailed elaboration of different components in the loss function.

\subsubsection{Physical Imaging Loss}
The physical imaging loss focuses on modeling the static 3D structure information in NeuralCMF according to the input echocardiogram videos. Because the physical imaging process of the 3DE varies from the 2DE, the formula for $\mathcal{L}$ varies accordingly:
\begin{equation}
\begin{aligned}
    &\mathcal{L}_{image}\\
    =&\left\{
    \begin{array}{ll}
        \sum_{t=0}^{T-1}\sum_{\vec{X}_{t}}\left\lVert O(\vec{X}_{t},t) - \mathbf{I}_{g}(\vec{X}_{t},t)\right\lVert_{2}^{2} & \text{3D\:\:Echo.} \\
        \sum_{t=0}^{T-1} \mathcal{L}_{CardiacField}\:(\text{Eqn.}\:\ref{eqn:loss_phy_loss}) & \text{2D\:\:Echo.}
    \end{array}
    \right.,
\label{eqn:img_loss}
\end{aligned}
\end{equation}
where $\mathbf{I}_{g}$ refers to the input 3DE or 2DE videos, $\Vec{X}_t$ is a 3D point at time $t$ and its definition domain is determined by the resolution of input videos, \textit{e.g.}, $\Vec{X}_t \in \{0,1,...,159\}^3$ for an input 3DE video with resolution $160^3$. When a 3DE video is given, the 3D structure of the heart can be supervised directly by comparing the network output $O(\Vec{X}_t,t)$ with the input video. However, when a 2DE video is given, the slicing operation (Eqn.~\ref{eqn:slicing_imaging}) in 2D echocardiography should be involved to generate a 2D image for comparison with the input 2D image. Note that the parameters for characterizing the angles of different 2D images are also optimized automatically in the training process. 

\subsubsection{Physical Motion Consistency Loss}
In echocardiogram videos, the intensity remains consistent in both the heart's tissues and blood-filled areas throughout the entire cardiac cycle \cite{de2012temporal}. Taking inspiration from this `consistent intensity' prior, we propose the physical motion consistency loss, ensuring the continuity of intensity for any given point within the 3D heart over time. Given a point $\vec{X}$ in the 3D heart at time $t$, \textit{i.e.}, $\vec{X}_t$, its corresponding point at time $t+1$ and $t-1$ can be calculated by,
\begin{equation}
\begin{aligned}
    \vec{X}_{t+1} &= \vec{X}_t + \vec{m}_{t \rightarrow t+1}(\vec{X}_t,t)\\
    \vec{X}_{t-1} &= \vec{X}_t + \vec{m}_{t \rightarrow t-1}(\vec{X}_t,t).
\label{eqn:fw_bw}
\end{aligned}
\end{equation}

Then, these new positions are fed into the $F_{\theta}^{Motion}$ to obtain the corresponding intensity values, forward and backward motions. Following the `consistent intensity' prior, the physical motion consistency loss could be mathematically modeled as follows,
\begin{equation}
\begin{aligned}
    &\mathcal{L}_{motion} \\
    =& \alpha_1\sum_{t = 0}^{T-1} \sum_{\vec{X}_t} \sum_{{i\in \{\pm 1\}}}(\left\lVert{O}(\vec{X}_{t+i},t+i) - {O}(\vec{X}_{t},t)\right\rVert^2_{2}\\
    + &{\alpha_2 \left\lVert \vec{X}_{t+i} + \vec{m}_{t \rightarrow t-i}(\vec{X}_{t+i},t+i) - \vec{X}_t\right\rVert^2_{2}}),
\label{eqn:motion_loss}
\end{aligned}
\end{equation}
where $T$ denotes the total frames in an entire cardiac cycle.

\subsubsection{Physical Cardiac Cycle Consistency Loss}
To enhance the precision of our model and align it with the physiological realities of cardiac function, we incorporate a physical cardiac cycle consistency loss. This loss reflects the fundamental cyclical behavior of cardiac motion throughout a single heart cycle, recognizing that the heart muscle returns to its initial shape after completing one cardiac cycle. The cycle consistency involves two primary constraints:
\begin{itemize}
\item \textbf{Intensity constraint.} The intensity of a particular point at the end of a cardiac cycle should closely align with its intensity at the beginning of the cycle. This intensity consistency condition reinforces the model's alignment with physical principles, ensuring the intensity is consistent across a full cycle.
\item \textbf{Motion constraint.} The heart's cyclic motion dictates that the cumulative forward or backward motions over the course of one cardiac cycle should sum to zero. This constraint ensures that the cardiac model adheres to the biological principle that the heart returns to its original shape after a complete cycle.
\end{itemize}
Mathematically, these two constraints can be formulated as,
\begin{equation}
\begin{aligned}
    \mathcal{L}_{cycle} &= {\alpha_1}\sum_{t=0}^{T-1}\sum_{\vec{X}_t}\left\lVert{O}(\vec{X}_{t+T},t+T) - {O}(\vec{X}_{t},t)\right\rVert^2_{2} \\
    &+ \alpha_2 \sum_{i\in\{\pm 1\}} \left\lVert\sum_{t=0}^{T-1}\vec{m}_{t\rightarrow t+i}(\vec{X}_{t},t)\right\rVert^2_{2}.
\label{eqn:cycle_loss}
\end{aligned}
\end{equation}

\subsubsection{Regularization Loss}
To reduce the noisy mismatch caused by the low signal-to-noise ratio in the input echocardiogram videos, two regularization terms are introduced in our model. 
The first term enforces consistency between the predicted forward and backward 3D motion vectors at each time step, ensuring that the forward motion vector $\vec{m}_{t \rightarrow t+1}(\vec{X}_t,t)$,  is congruent with the corresponding backward motion vector, $\vec{m}_{t \rightarrow t-1}(\vec{X}_t,t)$.Additionally, Eqns. \ref{eqn:motion_loss}, the physical motion consistency loss, complements this by aligning the $\vec{m}_{t \rightarrow t-1}(\vec{X}_t,t)$ and $\vec{m}_{t \rightarrow t+1}(\vec{X}_{t-1},t-1)$ to have equivalent magnitudes and opposite directions. Overall, these provisions ensure that the forward motion at each time step have the same magnitude and direction. This consistency contributes to smoother forward motion across consecutive time steps.
Concurrently, the purpose of the second term is to limit the magnitude of the 3D motion vectors at each time step, effectively preventing the occurrence of excessively large motions. By combining these two terms, the model effectively eliminates unpredictable and excessive movements, ensuring the motion is fluid, consistent, and smooth.
The regularization loss is thus formulated as follows,
\begin{equation} 
\begin{aligned}
    \mathcal{L}_{reg} = \sum_{t=0}^{T-1}\sum_{\Vec{X}_t}(\left\lVert \vec{m}_{t \rightarrow t+1}(\vec{X}_{t},t) + \vec{m}_{t \rightarrow t-1}(\vec{X}_{t},t) \right\rVert_{1}\\ +\left\lVert \vec{m}_{t \rightarrow t+1}(\vec{X}_{t},t) \right\rVert_{1} + \left\lVert \vec{m}_{t \rightarrow t-1}(\vec{X}_{t},t) \right\rVert_{1}).
\label{eqn:reg_loss}
\end{aligned}
\end{equation}

The algorithm presented in Algorithm \ref{full_algo} outlines the detailed pipeline of the proposed NeuralCMF. This algorithm aims to reconstruct the 3D structure and motion vectors from input multi-view 2DE or 3DE videos, all without requiring paired datasets. Initially, a batch of 4D coordinates is randomly selected and normalized (line 5). Subsequently, the intensity and motion vectors are obtained by inputting these coordinates into the respective neural networks (lines 6-9). Following this, the intensity and motion vectors are computed after moving forward and backward through one cardiac cycle (lines 10-14). In the fourth step, various loss functions are computed based on the definitions provided in Eqns. \ref{eqn:img_loss} through \ref{eqn:reg_loss} (lines 15). Finally, the combined loss function (line 16) supervises the update of the network parameters (line 17). These steps are iterated thousands of times. Upon convergence of training, the reconstructed motion vectors at any given space-time point can be obtained by inputting the coordinates into the network (line 6).

\begin{algorithm}
\caption{\textbf{Training Details of NeuralCMF}}
\label{full_algo}
\DontPrintSemicolon
\small
\KwIn{Spatial-temporal coordinates set $\{(\vec{X}_{t}, t)\}$. Multi-view 2DE/3DE video within one cardiac cycle.}
Initialize the parameters $\theta$ of $F_{\theta}^{motion}$.\;
Normalize coordinates within the range of $[0, 1]$.\;
\For{each training iteration}{
    Randomly select a batch of $(\vec{X}_t, t)$.\;
    Compute ${O}(\vec{X}_{t},t), {\vec{M}(\vec{X}_{t},t)}$ by Eqn. \ref{eqn:fw}.\;
    Compute $\vec{X}_{t+1}$ and $\vec{X}_{t-1}$ by Eqn. \ref{eqn:fw_bw}.\;
    Compute ${O}(\vec{X}_{t+1},t+1), {\vec{M}(\vec{X}_{t+1},t+1)}$ by Eqn. \ref{eqn:fw}.\;
    Compute ${O}(\vec{X}_{t-1},t-1), {\vec{M}(\vec{X}_{t-1},t-1)}$ by Eqn. \ref{eqn:fw}.\;
    \For{i from 1 to T}{
        Compute ${O}(\vec{X}_{t+i},t+i), {\vec{M}(\vec{X}_{t+i},t+i)}$ by Eqn. \ref{eqn:fw}.\;
        Compute ${O}(\vec{X}_{t-i},t-i), {\vec{M}(\vec{X}_{t-i},t-i)}$ by Eqn. \ref{eqn:fw}.\;
        Compute $\vec{X}_{(t+i)+1}$ and $\vec{X}_{(t-i)-1}$ by Eqn. \ref{eqn:fw_bw}.\;
        }  
    Compute each loss term using Eqns. \ref{eqn:img_loss}, \ref{eqn:motion_loss}, \ref{eqn:cycle_loss} and \ref{eqn:reg_loss}. \;
    $\mathcal{L} = \mathcal{L}_{image} + \mathcal{L}_{motion} + \mathcal{L}_{cycle} +\mathcal{L}_{reg}$.\;
    $\theta^{*} = \arg\min_{\theta} \mathcal{L}$.}
\end{algorithm}

\section{Experiments}
\subsection{Datasets}
We employ three distinct echocardiogram video datasets to evaluate the proposed NeuralCMF, each contributing unique characteristics and challenges.

\begin{figure*}[ht!]
\begin{center}
\includegraphics[width=\textwidth]{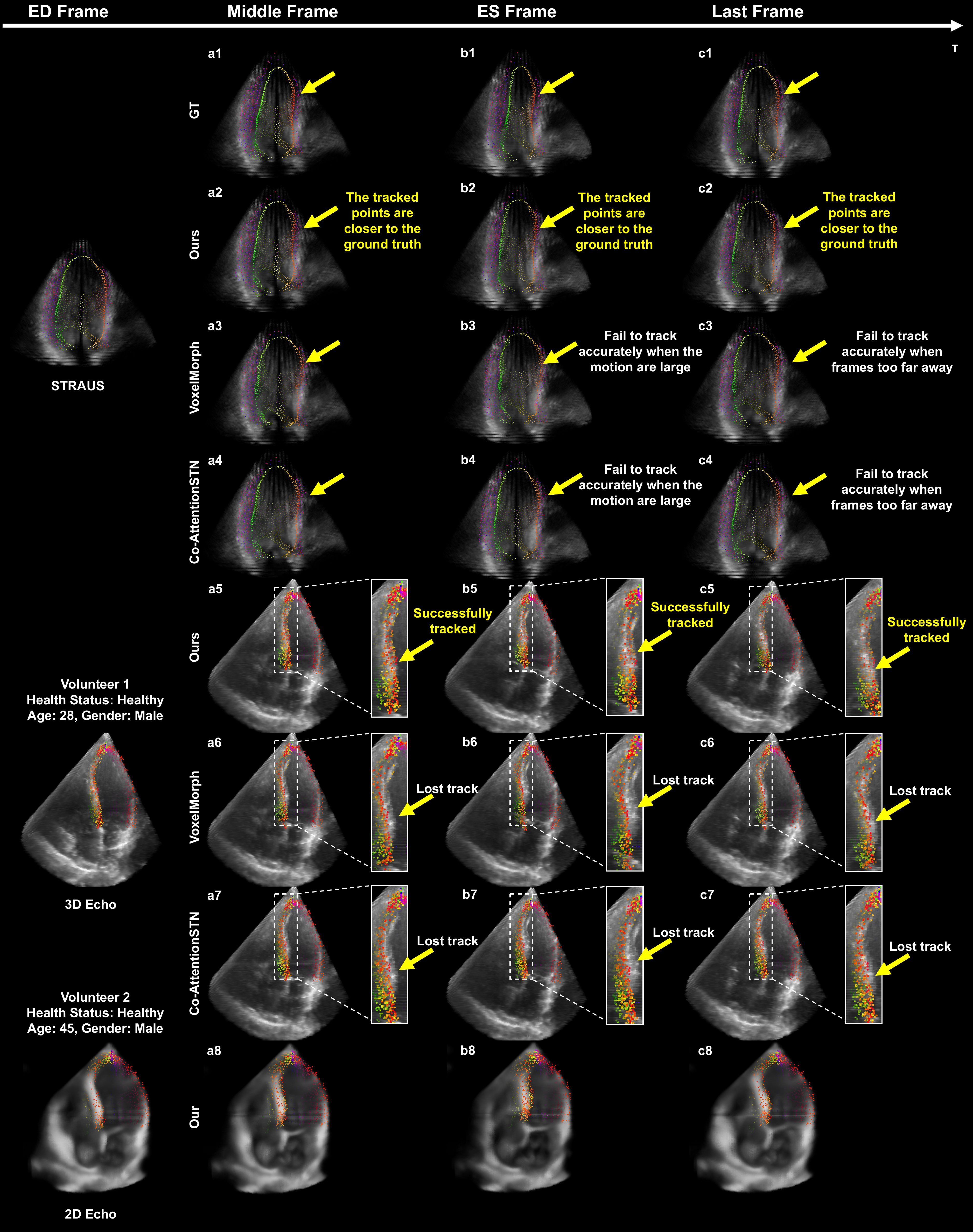}
\end{center}
\caption{This figure presents the qualitative results of point tracking using our method, VoxelMorph \cite{balakrishnan2019voxelmorph}, and Co-AttentionSTN \cite{ahn2023co} on the STRAUS \cite{alessandrini2015pipeline}, 2DE, and 3DE video datasets. The query points that need to be tracked during the ED phase are displayed in the first column. The following three columns visually demonstrate the movement of these points across the full cardiac cycle. In the 2DE video datasets, only our results are showcased since other methods are incompatible with multi-view 2DE datasets.}
\label{point_tracking}
\vspace{-2cm}
\end{figure*}

\subsubsection{STRAUS Datasets}
The open-source 3D strain assessment in ultrasound (STRAUS) dataset \cite{alessandrini2015pipeline} consists of 8 distinct volumetric sequences, each corresponding to a specific physiological condition. These include one healthy sequence, four ischemic cases, and three simulations of dilated cardiomyopathy. The ischemic cases are characterized by occlusions in various coronary arteries: the proximal or distal parts of the left anterior descending coronary artery (LADprox and LADdist, respectively), the left circumflex coronary artery (LCX), and the right coronary artery (RCA). The simulations of dilated cardiomyopathy consist of one case with a synchronous activation pattern (sync) and two cases of dyssynchrony due to left bundle branch block (LBBBsmall and LBBBlarge). These dyssynchronous cases are further distinguished by a progressively longer delay in activating the septum and lateral wall. Importantly, this dataset is accompanied by ground truth motion vectors, which have been derived from the movement of the left ventricular (LV) mesh. These vectors are instrumental in modeling the heart's motion, providing a valuable reference for analyzing and validating computational approaches. The 3D video in each data has a resolution $224\times 176\times 208$ with $34$ frames. This comprehensive dataset facilitates a robust examination of various heart conditions and serves as a valuable resource for analyzing and validating cardiac motion modeling.

\subsubsection{2DE and 3DE Video Datasets}
This dataset contains 127 video pairs from different volunteers, featuring both 2DE and corresponding 3DE videos for each individual.
The 2DE video dataset was acquired using commercial PHILIPS EPIQ 7C/IE ELITE machines with S5-1 2DE probes and SIEMENS ACUSON SC2000 PRIME machines with a 4V1c 2DE probe. During the imaging process, sonographers performed 360-degree rotations around the apex of the heart for each individual volunteer. Subsequently, the 2DE videos were synchronized based on concurrently recorded ECG signals. Any extraneous elements, such as text, electrocardiogram information, or other unrelated details outside the scanning sector, were cropped and masked.
The images were subsequently resized to $160\times 160$-pixel dimensions utilizing a bicubic interpolation filter and consist of 36 frames. In addition to the 2DE videos, the 3DE video dataset was acquired for each volunteer using commercial PHILIPS EPIQ 7C machines equipped with an X5-1 3DE probe.
These 3D datasets offer a resolution of $160\times 160\times 160$ pixels and contain 12 frames.
These 3D images were captured using a standard four-chamber apical view.

\subsection{Implementation Details}
The proposed NeuralCMF is implemented using the PyTorch framework. We use the FusedAdam optimizer, which is an implementation of the Adam optimizer \cite{kingma2014adam} that fuses multiple CUDA kernels and uses half-precision arithmetic to improve GPU memory usage and performance. The learning rate, initialized as 0.0001, is decayed according to the cosine annealing schedule. We trained our model for $10000$ iterations with 8192 points sampled in each iteration, and $\alpha_1 = 1$, $\alpha_2 =0.1$, $\alpha_3 = 0.01$. For every individual point, the position coordinate $\vec{X}$ is normalized to the range $\left[0,1\right]$. The time $t$ is subjected to repetition with the period of the cardiac cycle to align our implementation with the cyclic nature of the heartbeat,
\begin{equation} 
    t_{norm} = {\frac{t\mod T}{T}}.
\end{equation}
For both the STRAUS dataset and the 3DE video dataset, NeuralCMF is designed to take 3D video sequences as input.
We sample the dataset, extracting 4 images per cardiac cycle. With regard to the 2DE video datasets, NeuralCMF takes the 2D images as input, and our model processes 200 different view sequences. 

\subsection{Results}
\begin{table*}[!ht]

\begin{center}
\caption{Quantitative comparisons on STRAUS, 2DE, and 3DE video datasets. For both the 3DE and 2DE video datasets, the ground truth motion is unavailable, making it impossible to compute the MTE and Cosine similarity. In the 2DE video datasets, only our results are showcased since other methods are incompatible with multi-view 2DE datasets.}
\label{quantitative}
\begin{tabular}{l|c|c|c|c|c|c}
\toprule
\parbox{1.5cm}{\centering Datasets }            & Methods           & Median tracking error (mm) ↓& Cosine similarity ↑  & DICE ↑                 & HD (mm) ↓             & Jaccard index ↑       \\ \midrule
\multicolumn{1}{c|}{\multirow{3}{*}{\parbox{1.5cm}{\centering STRAUS \\ (LV Area)}}} & VoxelMorph  \cite{balakrishnan2019voxelmorph}      & $3.72 \pm 0.32$                & $0.68 \pm 0.04$          & $0.74 \pm 0.05$           & $3.08 \pm 0.12$           & $0.58 \pm 0.06$          \\
\multicolumn{1}{c|}{}                        & Co-AttentionSTN \cite{ahn2023co}   &            $4.41 \pm 0.48$                &         $0.69 \pm 0.08$             &           $0.71 \pm 0.04$           &       $3.30 \pm 0.10$               & $0.56 \pm 0.05$                     \\
\multicolumn{1}{c|}{}                        & Ours           &   $\textbf{2.55} \pm \textbf{0.07}$ & $\textbf{0.85} \pm \textbf{0.04}$ & $\textbf{0.82} \pm \textbf{0.02}$ & $\textbf{3.02} \pm \textbf{0.01}$ & $\textbf{0.69} \pm \textbf{0.03}$ \\ \midrule
\multirow{3}{*}{\parbox{1.5cm}{\centering 3D Echo \\ (LV Area)}}                     & VoxelMorph \cite{balakrishnan2019voxelmorph}       & -                          & -                    & $0.69 \pm 0.06$          & $3.11 \pm 0.14$          & $0.63 \pm 0.17$          \\
                                             & Co-AttentionSTN \cite{ahn2023co}  & -                          & -                    &             $0.65 \pm 0.10$         &        $3.20 \pm 0.23$              &        $0.61 \pm 0.09$              \\
                                             & Ours              & -                          & -                    & $\textbf{0.77} \pm \textbf{0.07}$ & $\textbf{3.04} \pm \textbf{0.05}$ & $\textbf{0.68} \pm \textbf{0.03}$           \\ \midrule
\parbox{1.5cm}{\centering 3D Echo \\ (LA Area)}               & \parbox{1.5cm}{\centering  \ \\ Ours}    & -                          & -                    & $\textbf{0.78} \pm \textbf{0.04}$          &$ \textbf{3.04} \pm \textbf{0.03}$           & $\textbf{0.68} \pm \textbf{0.05}$          \\ 
\parbox{1.5cm}{\centering (RA Area)}               & Ours              & -                          & -                    & $\textbf{0.75} \pm \textbf{0.06}$          &$ \textbf{3.20} \pm \textbf{0.19}$           & $\textbf{0.67} \pm \textbf{0.07}$          \\ \midrule
\parbox{1.5cm}{\centering 2D Echo \\ (LV Area)}               & \parbox{1.5cm}{\centering  \ \\ Ours}              & -                          & -                    & $\textbf{0.75} \pm \textbf{0.08}$          &$ \textbf{3.24} \pm \textbf{0.40}$           & $\textbf{0.67} \pm \textbf{0.04}$          \\
\parbox{1.5cm}{\centering (LA Area)}               & Ours              & -                          & -                    & $\textbf{0.75} \pm \textbf{0.05}$          &$ \textbf{3.20} \pm \textbf{0.12}$           & $\textbf{0.67} \pm \textbf{0.03}$          \\
\parbox{1.5cm}{\centering (RA Area)}               & Ours              & -                          & -                    & $\textbf{0.74} \pm \textbf{0.09}$          &$ \textbf{3.25} \pm \textbf{0.21}$           & $\textbf{0.67} \pm \textbf{0.09}$          \\\bottomrule
\end{tabular}
\end{center}
\end{table*}

\begin{figure*}[ht!]
\begin{center}
\includegraphics[width=0.87\textwidth]{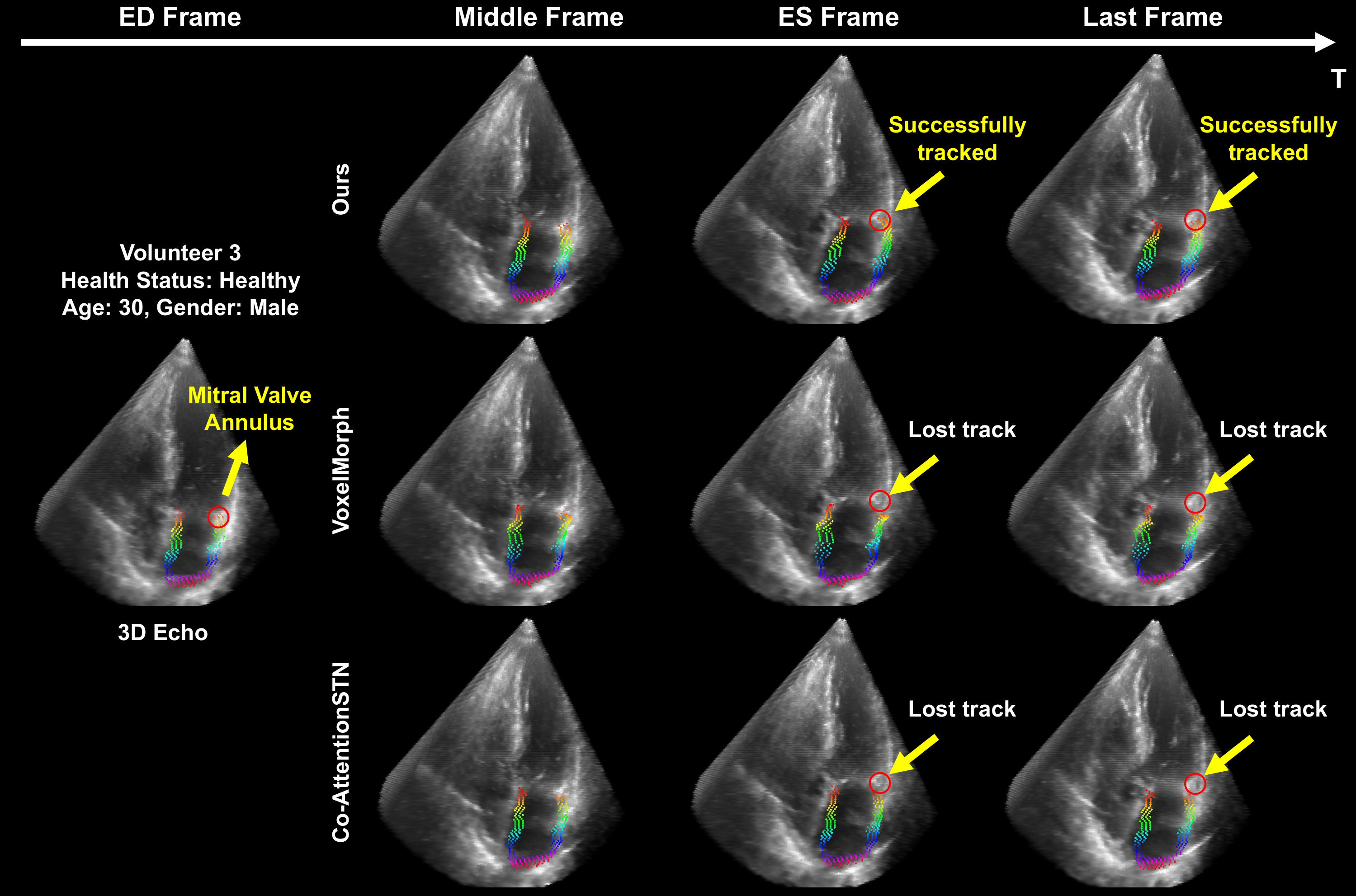}
\end{center}
\caption{This figure illustrates the tracking performance of our method, VoxelMorph, and Co-AttentionSTN for points on the left atrial (LA) myocardium within 3DE video datasets. The initial column highlights the reference points set for tracking during the ED phase. The subsequent three columns visually demonstrate the movement of these points across the full cardiac cycle. It is of significance to note that our method demonstrates superior accuracy in tracking points on the mitral valve annulus, a challenge not achieved by either VoxelMorph or Co-AttentionSTN.}
\label{track_LA}
\end{figure*}

\subsubsection{Evaluation Metrics}
The experimental investigations conduct on the STRAUS dataset are evaluated using the provided ground-truth motion vectors. This facilitates the calculation of two key metrics: the median tracking error (MTE) and the cosine similarity. The MTE metric, determined by the $\mathcal{L}_{2}$-norm between the ground truth and predicted motion vectors, evaluates the quantitative accuracy of the tracking, while the cosine similarity serves as a measure of the alignment of the directional components of the motion vectors, with 1 signifying perfect alignment and -1 indicating completely opposite directions. Together, these metrics provide a comprehensive evaluation of both the magnitude and direction of the computed motion fields.

In the case of the 2DE and 3DE video datasets \cite{shen2023cardiacfield}, where ground-truth motion vectors are unavailable, we rely on segmentation-based assessments. We employ metrics such as the hausdorff distance (HD), dice similarity Coefficient and jaccard index, comparing them against provided segmentation masks across two phases. The segmentation masks are labeled by an experienced echocardiographer. Using the learned motion vectors, we warp the 3D images from the end-diastolic (ED) phase to the end-systolic (ES) phase. Subsequently, we compare the myocardium segmentation of the LV in the four-chamber apical view from the warp ES phase to the myocardium segmentation in the real ES phase.

The NeuralCMF is evaluated against various state-of-the-art registration/motion tracking algorithms across each dataset for a robust quantitative comparison. Specific comparisons are made with other deep learning-based methods, including VoxelMorph \cite{balakrishnan2019voxelmorph} and Co-AttentionSTN \cite{ahn2023co}.  
The STRAUS dataset comprises 8 different 3DE sequences, we employ a 5/1/2 split strategy—utilizing 5 sequences for training, 1 for validation, and 2 for testing. 
Recognizing the constraints posed by the relatively small dataset, we have implemented a strategy to enhance the diversity and relevance of our training set, following the precedent established by \cite{ahn2023co}. Specifically, we select additional frame pairs from each sequence that exhibited significant variations in the cardiac cycle, indicated by a DICE overlap below 0.7. This method yields a robust dataset of 415 frame pairs for training and 83 for validation. This strategic selection ensures a more equitable comparison across different methods.
For our 3D echo datasets, we prepare 594 images for training, 91 for validation, and select sequences from 100 individuals for testing.  
It is important to note that unlike the above-mentioned other deep learning-based methods, NeuralCMF does not rely on a training dataset. We directly apply our method to the testing dataset to obtain motion results, which are then used for further comparison. This direct application streamlines the process, eliminating the need for a training phase and showcasing NeuralCMF's efficiency and adaptability in handling motion tracking tasks across datasets.

\begin{figure}[ht!]
\begin{center}
\includegraphics[width=\linewidth]{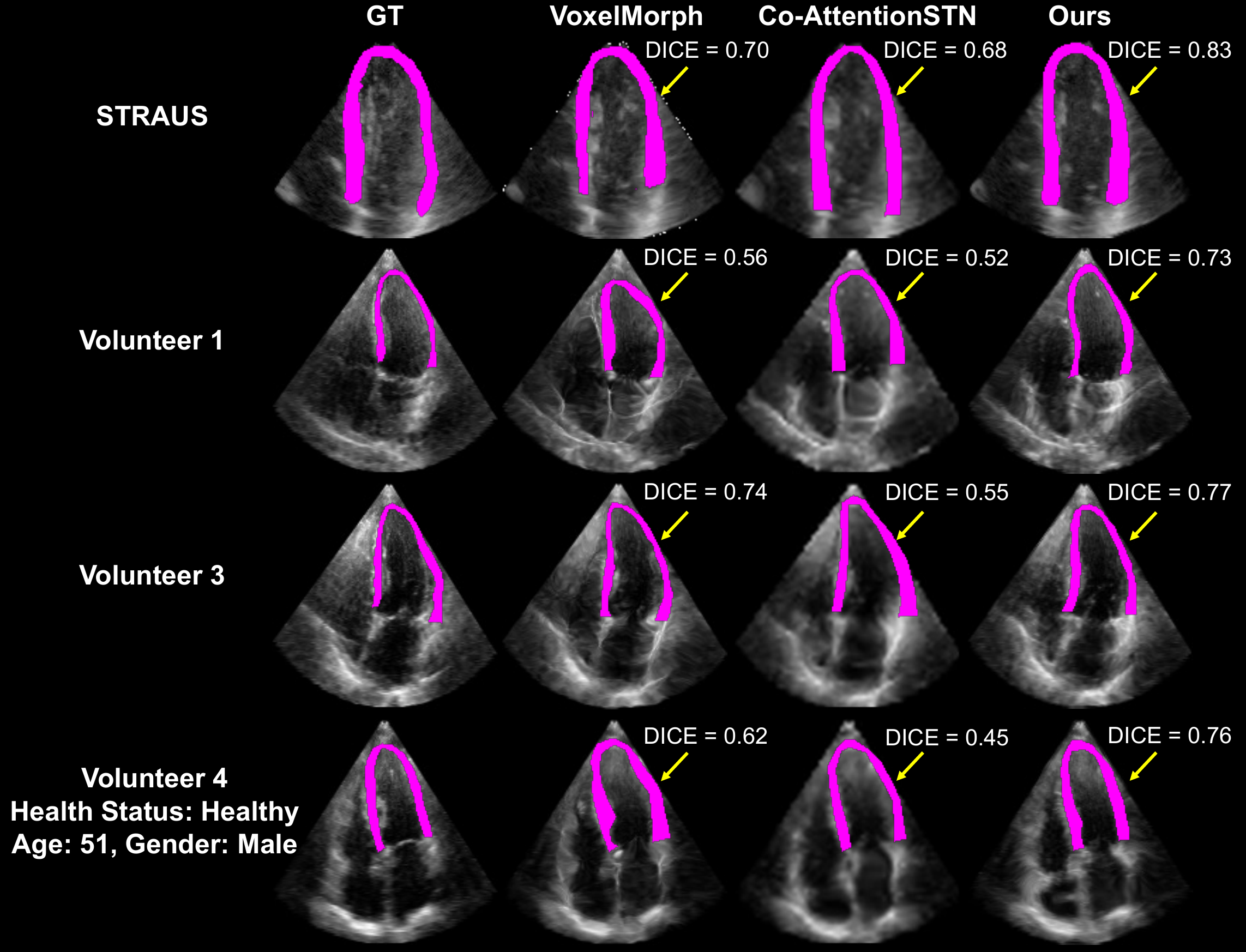}
\end{center}
\caption{This experiment demonstrates that our model offers a more precise motion estimation, with the warped image aligning closely to the actual physiological structure of the heart. The figure illustrates a comparison of our warped image results with those obtained using other deep-learning-based methods. The warping process involves taking a 3D image at the ED phase and transforming it with estimated displacements to create corresponding 3D images at the ES phase. Subsequently, we slice the apical four-chamber view for a detailed comparison. For clarity, the LV myocardium is segmented, and the DICE index is computed between the ground truth and warped ES phase images.}
\label{warp_results}
\end{figure}

\subsubsection{Results on STRAUS and 3DE Video Datasets}
In Table \ref{quantitative}, we offer quantitative comparisons of our proposed method using the STRAUS and the 3DE video datasets.
The STRAUS dataset is limited exclusively to the LV, necessitating a focused comparison on this region across all methods to ensure an equitable evaluation. Regarding the 3DE video datasets, the critical role of the LV in cardiac analyses within clinical settings directs standard 3D ultrasound data collection predominantly towards the LV. Although myocardial data for the LV, left atrium (LA), and right atrium (RA) are complete, the presence of ribs often obstructs ultrasound imaging, frequently resulting in incomplete myocardial data for the right ventricle (RV) \cite{ostenfeld2012manual,seo2020right}. This limitation impacts the effectiveness of myocardial motion tracking in the RV, prompting us to include experimental analyses for the LV, LA, and RA.

In the LV area, our method consistently outperforms competitors like VoxelMorph and Co-AttentionSTN across all evaluation metrics. It is noteworthy that the DICE coefficients for both the LA and RA regions exceed 0.7, indicating high accuracy in tracking. The results from the 3DE Video Datasets are lower than those from the STRAUS datasets. This discrepancy can be attributed to the higher signal-to-noise ratio present in the STRAUS datasets compared to the 3DE Video Datasets.
Figure \ref{point_tracking} illustrates the tracking results for both the STRAUS and 3D datasets. The images of LV myocardium at different frames are displayed from left to right. In the leftmost image, a total of 2500 points from the myocardium are selected and colored. Then, the corresponding points obtained by different motion tracking algorithms are plotted in the right 3 images. NeuralCMF demonstrates robust and consistent tracking capabilities throughout the cardiac cycle. Specifically, within the STRAUS datasets, tracking points transition from the ED to the ES phase (representing the largest movement in the cardiac cycle). There are two problems in tracking results of VoxelMorph and Co-AttentionSTN. First, the estimated motion displacement is shorter than the ground truth. The range of colored tracking points on the endocardium appears to be larger than the ground truth, as indicated by the yellow arrow in Fig. \ref{point_tracking}b1-b4. Second, points on the endocardium, which are uniformly sampled in the ED phase, become non-uniform in the ES phase as shown in Fig. \ref{point_tracking}b3, resulting in a loss tracking of starting points. In contrast, our method aligns myocardium points with the ground truth more precisely than these competing methods.

Regarding the 3DE video datasets, our approach consistently maintains tracking of the point as the myocardium transitions from ED to ES. Conversely, other methods frequently exhibit tracking inconsistencies, leading to point misplacement or loss during this significant stage. As illustrated in Fig. \ref{point_tracking}b6 and b7, both VoxelMorph and Co-AttentionSTN fail to track points on the interventricular septum (as indicated by the yellow arrow) during the ES phase. Particularly when myocardium movements are more substantial than in the ED phase, VoxelMorph and Co-AttentionSTN tend to lose track. Our method, however, reliably monitors these points across the entire cardiac cycle. Additionally, as shown in Fig. \ref{point_tracking}c5 - c7, our method outperforms VoxelMorph and Co-AttentionSTN in tracking myocardium movement, even over prolonged durations. This superiority is evident in the last frame of Fig. \ref{point_tracking}, where our method continues to track the interventricular septum accurately, whereas VoxelMorph and Co-AttentionSTN both lose their tracking points.

In addition to the LV area, we conducted comprehensive experimental analyses for the LA and RA, regions where data acquisition is typically more complete. To demonstrate the broad applicability and effectiveness of our method across different cardiac regions, we incorporated both quantitative (refer to Table \ref{quantitative}) and qualitative assessments (see Fig. \ref{track_LA} and Fig. \ref{More_tracking_results3d}).
For instance, as illustrated in Fig. \ref{track_LA}, we select key points on the LA myocardium, which are displayed in the first column. The subsequent three columns are present the tracking results of various methods during different phases. Notably, VoxelMorph's results become distorted during heart movement. In contrast, our method's tracking outcomes remain precise and consistent. It is pivotal to highlight the tracking of the points on the mitral valve annulus. NeuralCMF stably and consistently tracks its movement, while both VoxelMorph and Co-AttentionSTN fail to track the mitral valve annulus. This further elucidates that only our method can accurately capture the longitudinal movement of LA myocardium during the cardiac cycle.

The comparison in Fig. \ref{warp_results} indicates that our model produces results that align more precisely with the heart's actual structure. Figure \ref{warp_results} offers visual comparisons on the results of the warping process applied to 3D images. Specifically, the 3D images at the end-diastolic (ED) phase are transformed using the estimated motion vectors to create corresponding 3D images at the end-systolic (ES) phase. We then slice the apical four-chamber view from this 3D model for a detailed comparison. To offer a clearer comparison, we segment the myocardium of the LV and compute the DICE index between the ground truth and the warped images at the ES phase. Notably, the DICE score of our method surpasses that of other methods.

\subsubsection{Coronary Heart Disease Classification}
\begin{figure*}[ht!]
\begin{center}
\includegraphics[width=\linewidth]{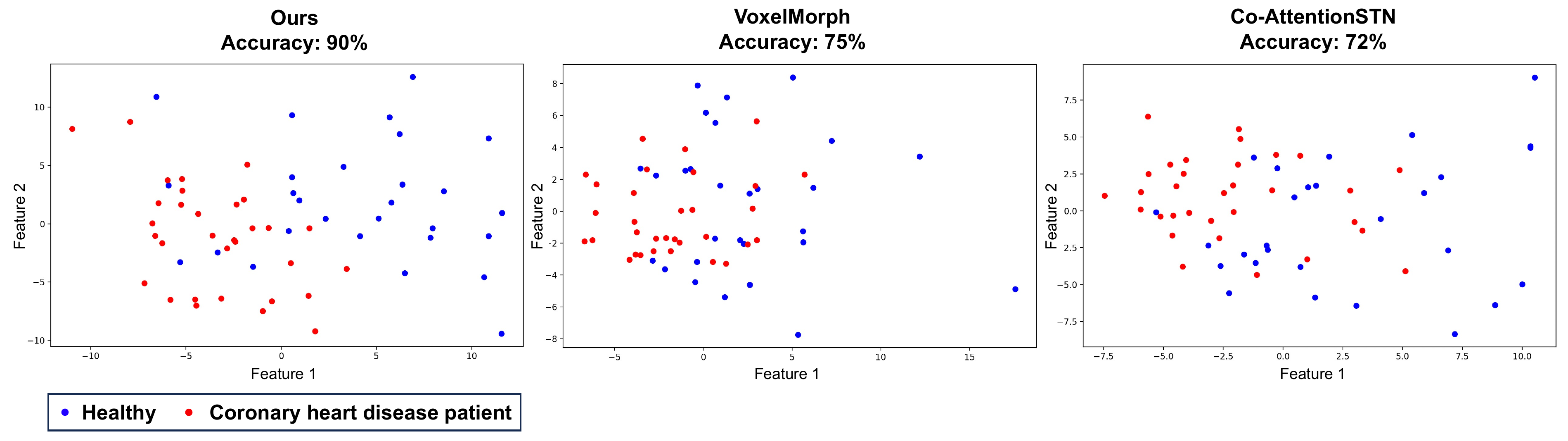}
\end{center}
\caption{Our method significantly enhances the clarity of CHD classification by projecting the motion vectors of test images onto a 2D PCA space. Our method achieves a 90\% classification accuracy using SVM with 10-fold cross-validation, while the accuracy of VoxelMorph is 75\% and Co-AttentionSTN is 72\%.}
\label{CHD_Classification}
\end{figure*}

To explore the additional value our approach brings to identifying disease pathologies, we conduct an experiment aimed at distinguishing between healthy individuals and patients with CHD using the motion data derived from our novel method. We extract 3D motion vectors from the LV area. Following the \cite{krebs2019learning}, these vectors are subsequently projected into a 10-dimensional (10-D) space using Principal Component Analysis (PCA), effectively capturing the core motion information. For the task of classification, we utilize the 10 most distinctive PCA components derived from the LV area. Our findings reveal that by using the motion information from the LV, we attain a 90\% accuracy rate in distinguishing healthy individuals from CHD patients, utilizing a 10-fold cross-validation method alongside Support Vector Machines (SVM). It's important to note that the methods we compared against, namely VoxelMorph \cite{balakrishnan2019voxelmorph} and Co-AttentionSTN \cite{ahn2023co}, represent the state of the art in registration/motion tracking algorithms, with classification accuracy rates of 75\% and 72\% respectively. The results are presented in Table \ref{CHD_Classification_tab}. The increased accuracy in our approach can be attributed to the more precise tracking of LV motion, which, as a result, enables more accurate classification of healthy individuals versus CHD patients.Additionally, we apply PCA to reduce the motion information of LV to a 2D vector for each subject, facilitating a visual representation of the separation between the two groups. This visualization is presented in Fig. \ref{CHD_Classification}.  This visualization elucidates the enhanced classification clarity provided by our method.This experiment highlights the capability of our methodology as an effective instrument for improving disease diagnostics, especially for CHD that may elude detection by existing approaches, e.g. chamber size and ejection fraction (EF).

\begin{table}[!ht]
\begin{center}
\caption{Comparison of CHD classification accuracy among different methods.}
\label{CHD_Classification_tab}
\begin{tabular}{c|c}
\toprule
Methods           &  Classification Accuracy            \\ \midrule
VoxelMorph  \cite{balakrishnan2019voxelmorph} &   75\%    \\
Co-AttentionSTN \cite{ahn2023co}   &          72\%        \\
Ours           & $\textbf{90\%}$  \\ \bottomrule
\end{tabular}
\end{center}
\end{table}

\subsubsection{Results on 2DE Video Datasets}
\begin{figure}[ht!]
\begin{center}
\includegraphics[width=\linewidth]{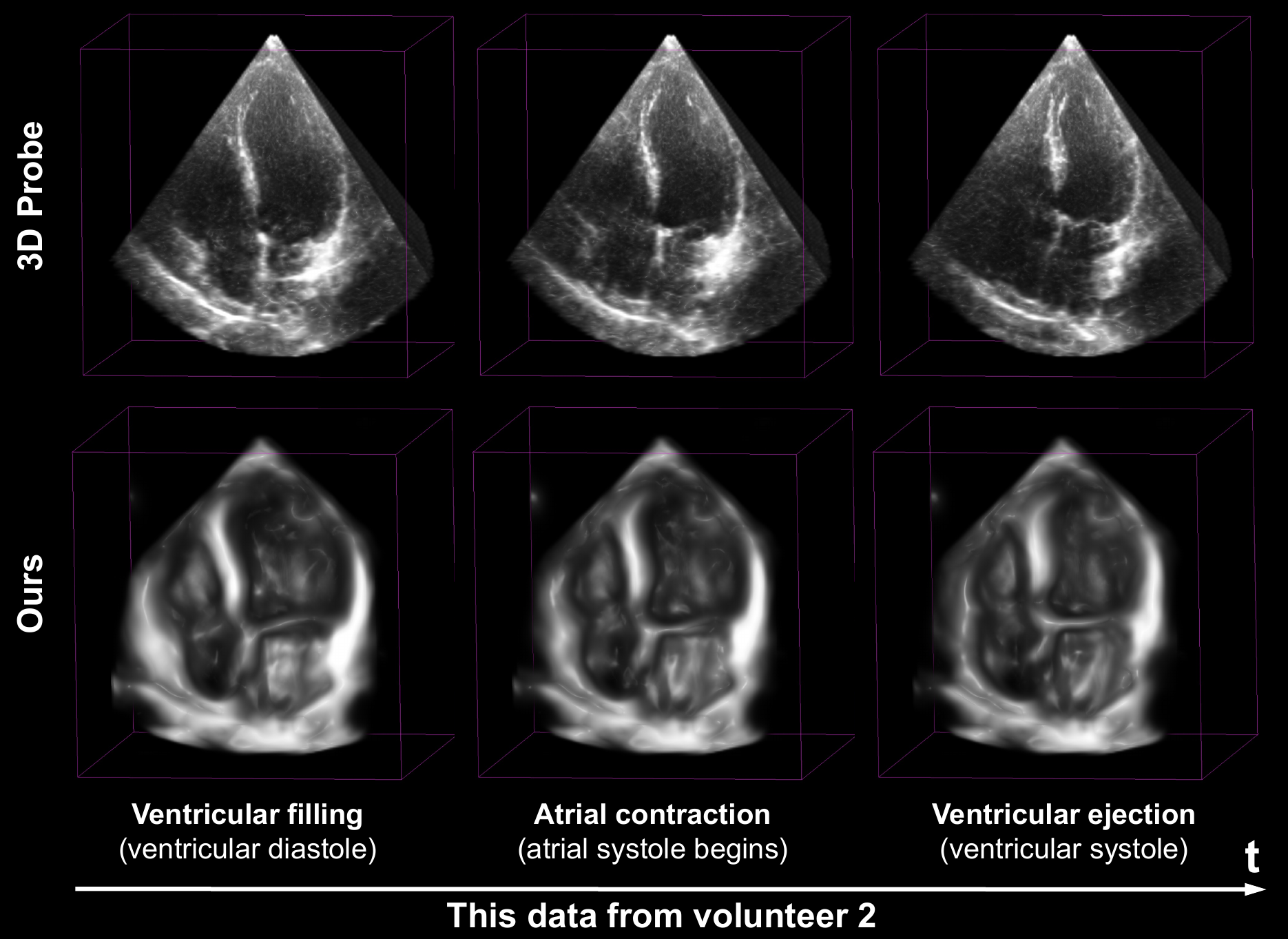}
\end{center}
\vspace{0.1cm}
\caption{We show snapshots of a reconstructed 3D dynamic heart within one cardiac cycle using our proposed method, and compare these with the outputs from images acquired via the 3DE probe.}
\label{3d_reconstruction_results}
\end{figure}

Our model transcends mere motion estimation by concurrently reconstructing a dynamic heart from 2DE videos, showcasing its capacity for handling multiple tasks simultaneously. The performance of our approach is validated through a comprehensive assessment that includes both quantitative and qualitative evaluations. The motion estimation results of the 2DE video datasets are thoroughly detailed in Table \ref{quantitative}, offering a statistical insight into the model's precision and consistency in tracking motion. Figure \ref{point_tracking} visually represents the point tracking results, enabling a more intuitive understanding of the method's effectiveness across cardiac cycles.
Additionally, our model's ability to reconstruct the 3D dynamic heart is evaluated by comparing its reconstruction results with the 3D hearts captured using a 3DE probe, as illustrated in Fig. \ref{3d_reconstruction_results}.
Compared to the 3DE probe, the 3D heart structure represented by our approach closely aligns with that of the 3DE probe. Moreover, our results exhibit a superior signal-to-noise ratio than the 3DE probe.
This experiment validates the multitasking ability of our method, demonstrating its robustness in handling complex cardiac imaging tasks.

\subsubsection{Ablation Study}
In this section, we explore the architecture of our network through a comprehensive ablation study, the findings of which are shown in Table \ref{ablation_network}. This analysis reveals that models incorporating static features outperform those relying solely on dynamic features. This may be attributed to the static feature captured by static SIREN, which can represent the cardiac structure that does not change over time, thereby providing a consistent reference frame. Such a reference is crucial for the dynamic SIREN, as it facilitates the differentiation of stationary areas from the areas exhibiting motion, enhancing the capability of the model to capture motion dynamics effectively. Furthermore, we conduct an experiment by substituting SIREN with a combination of positional encoding (PE) and MLP. The results presented in Table \ref{ablation_network} reveal a notable decrease in performance upon substituting SIREN with PE + MLP. This finding highlights the superior efficacy of the SIREN network compared to the PE + MLP in our architecture. Our decision to model forward and backward motions separately is not solely based on empirical evidence; it also holds significant physical relevance. This differentiation allows us to ensure that the color of any given point in the current frame remains consistent with both its preceding and succeeding frames, thereby preserving continuity across frames. To substantiate this approach, we conduct additional experiments comparing the performance of a model that only includes forward motion against our method, which treats forward and backward motions distinctly. The findings, presented in Table \ref{ablation_network}, demonstrate a significant degradation in performance when backward motion is merely the inverse of forward motion. Such a decline is attributable to the model's oversight of backward motion, wherein a point's color in the current frame is aligned only with the succeeding frame, neglecting the color continuity with the preceding frame.
\begin{table}[!t]
\begin{center}
\caption{Ablation Study on Network Structure.}
\label{ablation_network}
\begin{tabular}{c|c|c}
\toprule
Network Structure           &  MTE & Cosine Similarity                  \\ \midrule
Only Dynamic Siren                     &   $3.07 \pm 0.26$    &     $0.76 \pm 0.03 $     \\
PE + MLP            &          $31.08 \pm 4.19$          &          $0.33 \pm 0.09$ \\
Only forward motion  &          $4.93 \pm 0.95$          &          $0.54 \pm 0.06$ \\
Ours           & $\textbf{2.55} \pm \textbf{0.07}$  & $\textbf{0.85} \pm \textbf{0.04}$\\ \bottomrule
\end{tabular}
\end{center}
\end{table}

To deeply understand each component within our loss function, we conduct a series of ablation studies. These studies analyze the influence of different components, with the performance metrics present in Table  \ref{ablation_loss}. Our analysis reveals that the motion cannot be tracked without physical motion consistency loss $\mathcal{L}_{motion}$ and physical cardiac cycle consistency loss $\mathcal{L}_{cycle}$. This emphasizes the crucial importance of physical loss we proposed in stabilizing the learning process and ensuring convergence to a solution. Additionally, the results from Table \ref{ablation_loss} provide compelling evidence for the significance of both coordinate and intensity constraints within our loss function. When the coordinate constraint is removed (indicated as wo/coordinate), we observe a drastic increase in MTE and a decrease in cosine similarity, underscoring the importance of spatial constraints in accurate motion tracking. Similarly, the removal of the intensity constraint (indicated as wo/intensity) results in significant degradation of performance, highlighting the necessity of intensity information for precise motion estimation. 

\begin{figure}
\begin{center}
\includegraphics[width=\linewidth]{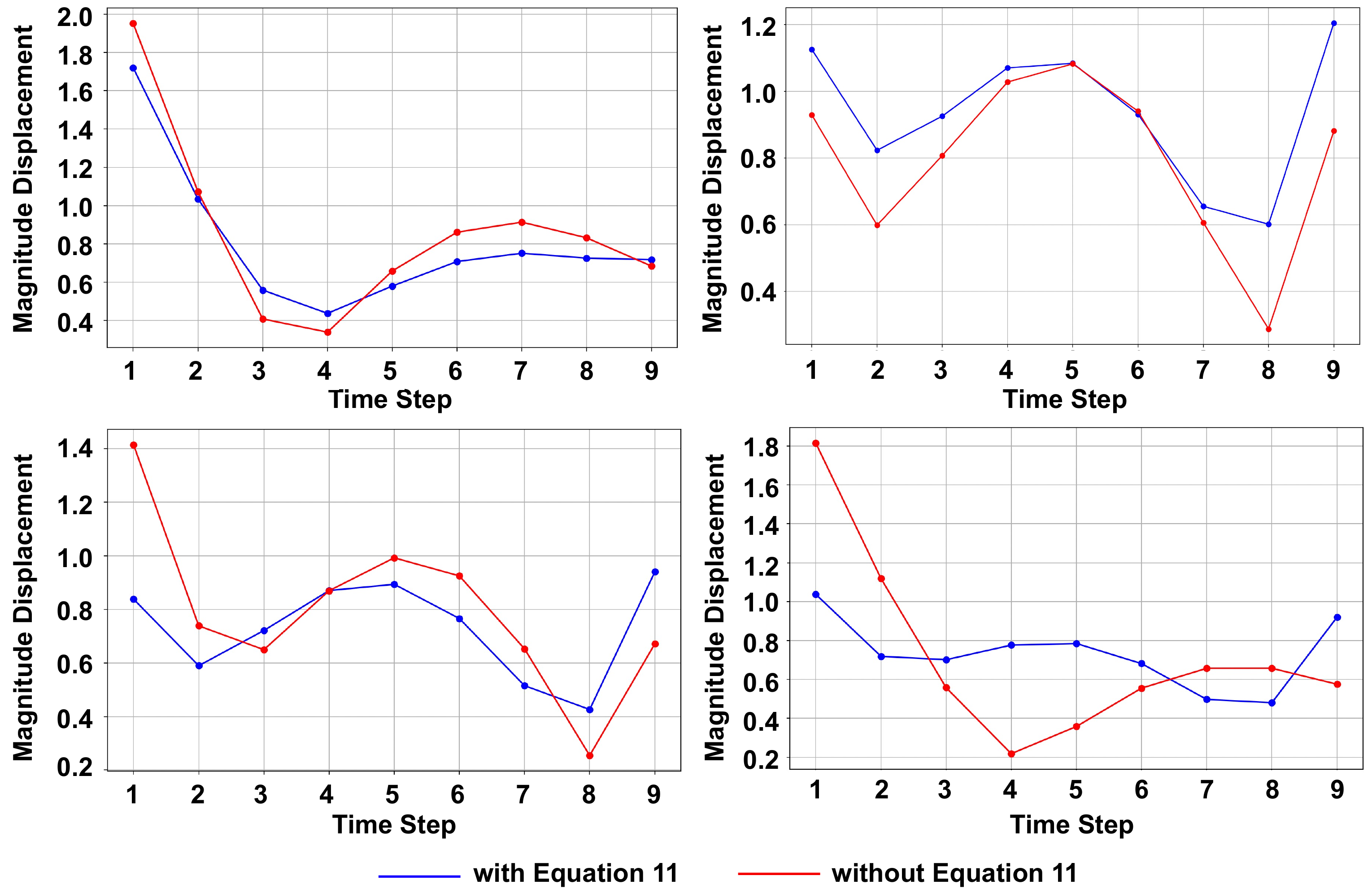}
\end{center}
\caption{The magnitudes of displacement between consecutive time steps for two scenarios: the blue line represents the model with $\mathcal{L}_{reg}$, and the red line represents the model without it. Each point reflects the displacement from one time step to the next. Lines with larger slopes indicate rapid changes in displacement, suggesting abrupt or significant motion changes and reduced smoothness. The gradual slopes and reduced fluctuations in the blue line demonstrate the smoothness introduced by $\mathcal{L}_{reg}$, compared to the sharper peaks and abrupt transitions in the red line. This comparison highlights the regularization strategy's effectiveness in enhancing cardiac motion smoothness.}
\label{smooth}
\end{figure}

To substantiate the effectiveness of $\mathcal{L}_{reg}$ term in enhancing motion smoothness, we conduct an experiment visualizing displacement changes between consecutive time points. We calculate displacement vectors between successive time steps using our model, both with and without the implementation of $\mathcal{L}_{reg}$. The magnitudes of these vectors are then plotted over time, as shown in Fig. \ref{smooth}. The graph illustrates the displacement magnitudes, where each point on the blue line (model with $\mathcal{L}_{reg}$) and the red line (model without $\mathcal{L}_{reg}$) represents the displacement from one time step to the next. A smoother curve, as exhibited by the blue line, indicates fewer abrupt changes, suggesting enhanced motion smoothness. In contrast, sharper peaks or significant fluctuations on the red line denote more abrupt changes, indicative of less smooth motion. The results confirm that the model incorporating $\mathcal{L}_{reg}$ achieves a smoother trajectory, as evidenced by the more gradual fluctuations and reduced slope steepness in the blue line compared to the red line. Furthermore, the ablation study present in Table \ref{ablation_loss} corroborates that the regularization term not only enhances the visual smoothness but also improves the accuracy of motion tracking.

\begin{table*}[!t]
\begin{center}
\vspace{0.15cm}
\caption{Systematic Ablation Study on Loss Function Components.}
\label{ablation_loss}
\begin{tabular}{c|c|c}
\toprule
Loss Function           &  MTE & Cosine Similarity                  \\ \midrule
$\mathcal{L}_{image}$                   &   $106.44 \pm 42.87$    &     $0.30 \pm 0.01 $     \\
$\mathcal{L}_{image}+\mathcal{L}_{motion}$  &   $123.18 \pm 23.46$    &   $0.31 \pm 0.20$           \\
$\mathcal{L}_{image}+\mathcal{L}_{motion}+\mathcal{L}_{cycle}$  & $2.93 \pm 0.14$ & $0.75 \pm 0.05$\\ 
$\mathcal{L}_{image}+\mathcal{L}_{motion}+\mathcal{L}_{cycle}+\mathcal{L}_{reg}$  & $\textbf{2.55} \pm \textbf{0.07}$ & $\textbf{0.85} \pm \textbf{0.04}$\\ 
$\mathcal{L}_{image}+\mathcal{L}_{motion}(w/o \ coordinate)+\mathcal{L}_{cycle}(w/o \ coordinate)+\mathcal{L}_{reg}$  & $128.61 \pm 87.62$ & $0.05 \pm 0.06$ \\
$\mathcal{L}_{image}+\mathcal{L}_{motion}(w/o \ intensity)+\mathcal{L}_{cycle}(w/o \ intensity)+\mathcal{L}_{reg}$       &   $5.75 \pm 0.87$    &     $0.15 \pm 0.10 $     \\ \bottomrule
\end{tabular}
\end{center}
\end{table*}

In tuning the hyperparameters for each component of our loss function, we adopt a systematic approach. Our loss function is composed of three parts: intensity loss with hyperparameter $\alpha_1$, coordinate loss with hyperparameter $\alpha_2$, and regularization loss with hyperparameter $\alpha_3$. Initially, we set all hyperparameters to 1 as a baseline. To understand the influence of each hyperparameter on motion estimation, we vary them from 0.01 to 2. The results of this hyperparameter tuning are presented in Table \ref{ablation_loss_weight_a1}, \ref{ablation_loss_weight_a2} and \ref{ablation_loss_weight_a3}. After careful evaluation of the hyperparameters' impact on our model's performance, we observe the following trends from three tables. For the intensity loss, the hyperparameter ($\alpha_1$) shows the optimal results at 1. The coordinate loss hyperparameter ($\alpha_2$) shows optimal results at 0.1. For the regularization loss hyperparameter ($\alpha_3$), a value of 0.01 yields the best performance. Consequently, we adopt the hyperparameter settings of 1 for intensity, 0.1 for coordinate, and 0.01 for regularization as our final model configuration, denoted as 'Ours' in the table. This combination results in the most effective performance, achieving an optimal balance across all metrics.

\begin{table}[!t]
\begin{center}
\caption{Hyperparameter Sensitivity Analysis of $\alpha_1$.}
\label{ablation_loss_weight_a1}
\begin{tabular}{c|c|c}
\toprule
\multicolumn{1}{c|}{\multirow{2}{*}{Hyper Parameters}} &\multicolumn{2}{c}{Intensity ($\alpha_1$)} \\ \cmidrule{2-3}
       & MTE     & Cosine Similarity    \\ \midrule
  0.01 & $5.74 \pm 0.89$ & $0.13 \pm 0.01$ \\
  0.1 & $5.75 \pm 0.85$ & $0.16 \pm 0.12$ \\ 
  1 & $\textbf{5.64} \pm \textbf{0.89}$ & $\textbf{0.32} \pm \textbf{0.03}$\\ 
  2 & $5.67 \pm 0.80$ & $0.29 \pm 0.07$ \\ \midrule
  Ours & $\textbf{2.55} \pm \textbf{0.07}$ & $\textbf{0.85} \pm \textbf{0.04}$\\ \bottomrule
\end{tabular}
\end{center}
\end{table}

\begin{table}[!t]
\begin{center}
\caption{Hyperparameter Sensitivity Analysis of $\alpha_2$.}
\label{ablation_loss_weight_a2}
\begin{tabular}{c|c|c}
\toprule
\multicolumn{1}{c|}{\multirow{2}{*}{Hyper Parameters}} &\multicolumn{2}{c}{Coordinate ($\alpha_2$)} \\ \cmidrule{2-3}
       & MTE     & Cosine Similarity      \\ \midrule
  0.01 & $5.64 \pm 0.81$ & $0.46 \pm 0.13$ \\
  0.1 & $\textbf{5.60} \pm \textbf{0.88}$ & $\textbf{0.47} \pm \textbf{0.09}$\\ 
  1 & $5.64 \pm 0.89$ & $0.32 \pm 0.03$ \\ 
  2 &  $5.75 \pm 0.85$ & $0.30 \pm 0.22$ \\ \midrule
  Ours & $\textbf{2.55} \pm \textbf{0.07}$ & $\textbf{0.85} \pm \textbf{0.04}$\\ \bottomrule
\end{tabular}
\end{center}
\end{table}

\begin{table}[!t]
\begin{center}
\caption{Hyperparameter Sensitivity Analysis of $\alpha_3$.}
\label{ablation_loss_weight_a3}
\begin{tabular}{c|c|c}
\toprule
\multicolumn{1}{c|}{\multirow{2}{*}{Hyper Parameters}} &\multicolumn{2}{c}{Regularization ($\alpha_3$)} \\ \cmidrule{2-3}
       & MTE     & Cosine Similarity       \\ \midrule
  0.01 & $\textbf{3.26} \pm \textbf{0.25}$ & $\textbf{0.78} \pm \textbf{0.04} $\\
  0.1 & $3.98 \pm 0.15$ &  $0.78 \pm 0.10$\\ 
  1 &  $5.64 \pm 0.89$ & $0.32 \pm 0.03$ \\ 
  2 &  $5.78 \pm 0.87$ & $0.07 \pm 0.07 $\\ \midrule
  Ours & $\textbf{2.55} \pm \textbf{0.07}$ & $\textbf{0.85} \pm \textbf{0.04}$\\ \bottomrule
\end{tabular}
\end{center}
\end{table}

We conduct an ablation study to examine how varying the percentage of observed ground truth (GT) pixels influences the performance of our method. We systematically vary the amount of GT pixels used to supervise the network training and evaluate the impact on performance metrics. The results are displayed in Table \ref{ablation_gt_number} and provide insightful trends. This table provides information on the impact of varying proportions (20\%, 40\%, 60\%, 80\%, and 100\%) of GT pixels on MTE and cosine similarity. Data is presented as mean $\pm$ standard deviation. Analyzing the content of the table, we can observe trends from two dimensions: MTE and cosine similarity. Overall, 100\% GT supervision yields the best results, evidenced by the decrease in MTE and the increase in cosine similarity.

\begin{table}[!t]
\begin{center}
\caption{Ablation Study on the Percentage of GT Pixels.}
\label{ablation_gt_number}
\begin{tabular}{c|c|c}
\toprule
Percentage of GT Pixels  &  MTE &Cosine Similarity      \\ \midrule
20\% &  $2.94 \pm 0.19$  &  $0.79 \pm 0.04 $   \\
40\% &  $2.83 \pm 0.03 $  &$0.80 \pm 0.06$ \\
60\% &   $2.81 \pm 0.24 $  &  $0.80 \pm 0.03 $   \\
80\% &   $2.75 \pm 0.02 $  &  $0.82 \pm 0.05 $   \\
100\% & $\textbf{2.55} \pm \textbf{0.07}$ &$\textbf{0.85} \pm \textbf{0.04}$\\ \bottomrule
\end{tabular}
\end{center}
\end{table}
\begin{figure*}[ht!]
\begin{center}
\includegraphics[width=\linewidth]{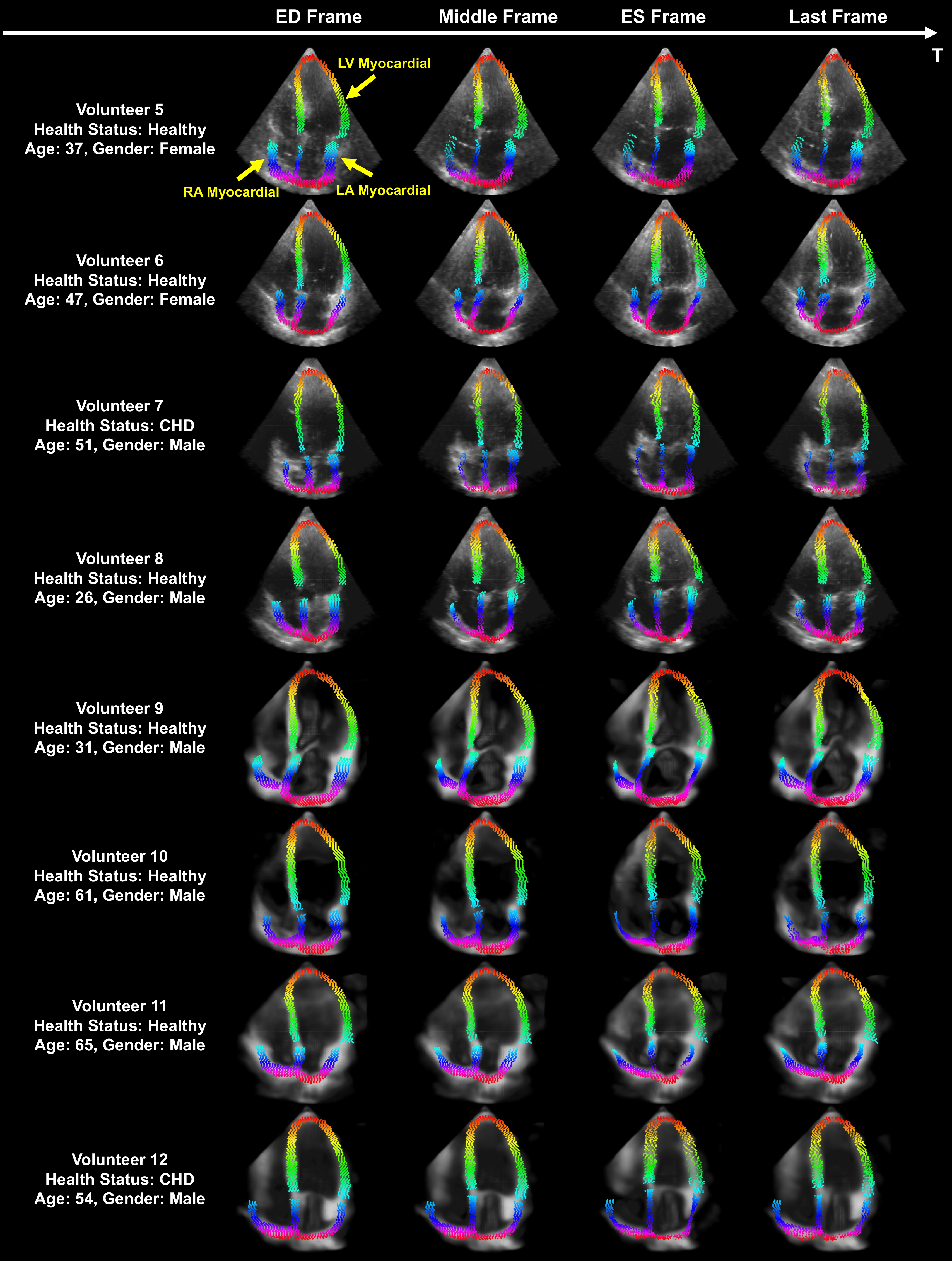}
\end{center}
\caption{Additional visual results of myocardial motion tracking for the LV, LA, and right atrium (RA) (as indicated by the yellow arrow) across diverse individuals in 2DE/3DE video datasets: data of volunteers 5-8 from 3D datasets and 9-12 from multi-view 2DE datasets. Additional results are available on our project's GitHub repository.}
\label{More_tracking_results3d}
\vspace{-2cm}
\end{figure*}

\begin{table}[!t]
\begin{center}
\caption{Comparison of running times (in minutes) for motion estimation of an individual on the STRAUS dataset.}
\label{running_time}
\begin{tabular}{c|c|c}
\toprule
Methods           &  Pre-Training & Motion Estimation                  \\ \midrule
VoxelMorph  \cite{balakrishnan2019voxelmorph}                     &    54.55     &     0.24      \\
Co-AttentionSTN \cite{ahn2023co}             &          94.13           &          $\textbf{0.08}$                                     \\
Ours (1 GPU)      &   $\textbf{0}$  &  8.58 \\ 
Ours  (4 GPUs)    &   $\textbf{0}$  &  3.83 \\\bottomrule
\end{tabular}
\end{center}
\end{table}

\begin{figure}
\begin{center}
\includegraphics[width=\linewidth]{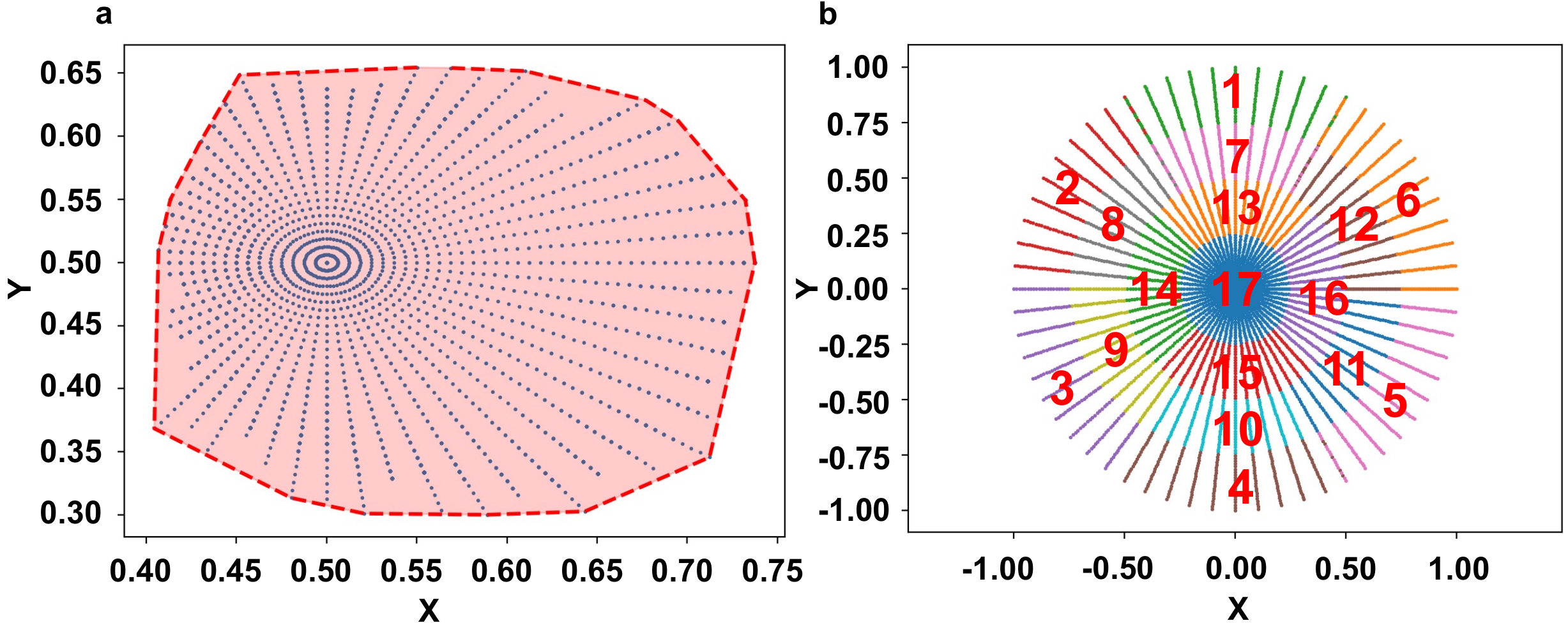}
\end{center}
\caption{The steps to transfer a point on the LV endocardial surface to the 17 American Heart Association (AHA) segments. (a) The process begins with mapping the points on the endocardial surface onto the short-axis plane. (b) Next, we correct this mapping by aligning the apical point to the center and arranging the points uniformly on a disk with a radius of 1. Lastly, we partition the points on the resulting disk in accordance with the AHA segment's specific disk definition.}
\label{AHA_Segments}
\end{figure}

\begin{figure*}
\begin{center}
\includegraphics[width=\linewidth]{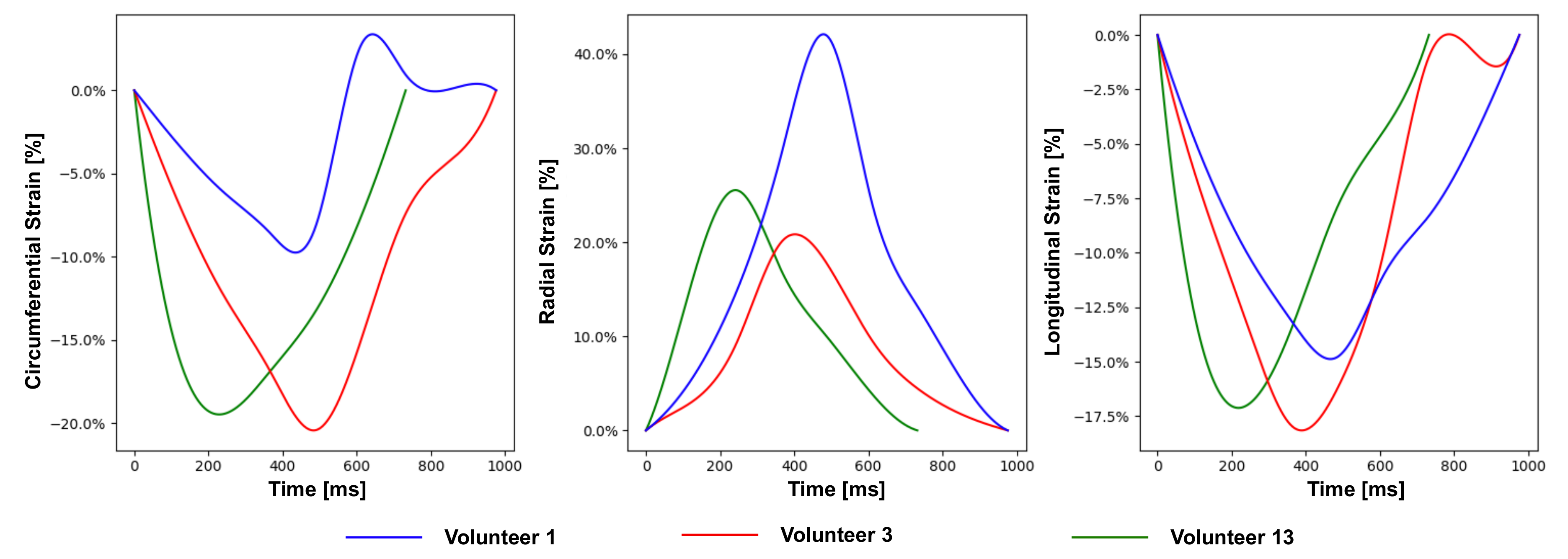}
\end{center}
\caption{The circumferential, radial, and longitudinal strain curves across the complete cardiac cycle for three volunteers.}
\label{strain_curve}
\end{figure*}

\subsection{Results on Myocardial Strain Computation}
Myocardial strain is a critical parameter in heart biomechanics and can be computed using Lagrangian strain \cite{de2012temporal}. The process of deriving Lagrangian strain begins with defining the deformation gradient tensor, denoted as $\mathbf{F}$, to characterize the displacement of points within the heart during deformation. The expression for the heart deformation gradient tensor $\mathbf{F}$ is given by,
\begin{equation} 
    \mathbf{F} = \mathbf{1} + \nabla \vec{\mathbf{m}}_{t \rightarrow t+1}(\vec{X}_{t},t),
\end{equation}
where $\mathbf{1}$ represents the identity tensor, and $\nabla \vec{\mathbf{m}}_{t \rightarrow t+1}$ is the forward motion field gradient. Our method has significant advantages for calculating the forward motion field gradient because the network's input is the coordinate and the output is the motion. This setup allows for the straightforward calculation of the motion field gradient using the neural network's automatic differentiation mechanism. This gradient can be further expressed as,
\begin{equation}
\nabla \vec{\mathbf{m}}_{t \rightarrow t+1}(\vec{X}_{t},t) = \begin{bmatrix}
\frac{\partial m_x}{\partial x} & \frac{\partial m_x}{\partial y} & \frac{\partial m_x}{\partial z} \\
\frac{\partial m_y}{\partial x} & \frac{\partial m_y}{\partial y} & \frac{\partial m_y}{\partial z} \\
\frac{\partial m_z}{\partial x} & \frac{\partial m_z}{\partial y} & \frac{\partial m_z}{\partial z}
\end{bmatrix},
\end{equation}
where the components $m_x$, $m_y$, and $m_z$ represent the components of the forward motion $\vec{{m}}_{t \rightarrow t+1}(\vec{X}_{t},t)$ in the $x$, $y$, and $z$ directions, respectively. Next, we arrive at the Lagrangian strain tensor $\mathbf{E}$, computed at point $\vec{X}_{t}$ at time $t$,
\begin{equation} 
    \mathbf{E}(\vec{X}_{t},t) = \frac{1}{2} (\mathbf{F^{\top}} \mathbf{F} - \mathbf{I}).
\end{equation}
To calculate the strain in a specific direction $\vec{d}$ of the LV, we project the strain in that direction using,
\begin{equation} 
    \mathbf{E}_{\vec{d}}(\vec{X}_{t},t) =\vec{d}^{\top} \mathbf{E}(\vec{X}_{t},t) \vec{d},
\end{equation}
where $\vec{d}$ is a $3 \times 1$ directional vector.

In our implementation, the longitudinal direction is $\vec{d}_l = \left[0,0,1 \right]^T$. To compute the radial direction $\vec{d}_r$, we first determine the normalized direction $\vec{d}_e$ of each point in the LV myocardium to its center, then deduce the radial direction as $\vec{d}_r = \vec{d}_e-(\vec{d}_e \cdot \vec{d}_l)\vec{d}_l$. Lastly, the circumferential direction $\vec{d}_c$ is computed as the cross product of the previous two vectors, \textit{i.e.}, $\vec{d}_c = \vec{d}_l \times \vec{d}_r$.

In the study of myocardial mechanics, a precise visualization of strain across different regions of the heart is essential. To this end, we present a method to visualize the strain across 17 American Heart Association (AHA) segments. This approach comprises the following steps:
\begin{enumerate}
\item Sampling Points: First, we sample points on the LV endocardial surface and extract their 3D coordinates. To achieve this, we uniformly slice the 3D heart into 30 2D images at the end-diastolic (ED) phase. These slices are performed along the long axis of the 3D heart.
\item Segmentation of Endocardial Area: Subsequently, we segment the endocardial area of the LV in each slice and determine the 3D position of that region.
\item Classification into AHA Segments: Following the approach proposed by \cite{de2012temporal}, we classify the points on the endocardial surface into 17 AHA segments. The specific steps shown in Fig. \ref{AHA_Segments} include: a). Mapping the points on the endocardial surface onto the short-axis plane. b). Correcting the mapping by aligning the apical point to the center. c). Arranging the points uniformly on a disk with a radius of 1. d). Partitioning the points on the resulting disk according to the disk definition of the AHA segment. e). Obtaining the 3D coordinates of the points in each AHA region on the disk.
\item Visualizing Strain Curves: We visualize strain curves for each anatomical direction. We accomplish this by averaging the strain values of all points in each AHA segment, thereby obtaining the global strain in that segment. The results are then plotted as a function of time.
\end{enumerate}
This analysis is performed on a subset of three patients. The results of the global strain curve in three directions are shown in Fig. \ref{strain_curve}.

\subsection{Complexity}
Table \ref{running_time} presents the running time of our method in comparison to VoxelMorph and Co-AttentionSTN using a single Nvidia A100 GPU with 40GB memory. Despite the longer training periods often associated with supervised learning techniques, they tend to offer shorter inference times per subject. In contrast, our approach may result in longer per-subject testing times.

Our model training involves 10,000 iterations, with 8,192 points sampled in each iteration. To optimize processing efficiency, we employ distributed data parallel (DDP) to distribute these 8,192 points across multiple GPUs. This strategy allows each GPU to process a smaller portion of the total points simultaneously, significantly reducing the processing time required for each training step. We conduct an experiment to train our model using four NVIDIA 3090 GPUs. As shown in Table \ref{running_time}, this approach reduced the motion estimation time to 3.83 minutes.

\section{Discussion and Conclusion}
The diagnostic, prognostic, and preventive measures associated with CVDs remain essential clinical concerns. 
Advances in cardiac motion tracking technologies, such as CMR feature tracking and speckle-tracking echocardiography have made deformation imaging a viable, reliable, and valuable clinical tool. Global or segmented strain measurement can objectively quantify myocardial deformation and provide a novel method for characterizing myocardial function, which is replacing the conventional heart function evaluation parameters. Given the ubiquity of echocardiography as a bedside imaging modality, its precision in monitoring myocardial movement is vital for identifying abnormalities in the heart.
In this context, we introduce the Neural Cardiac Motion Field (NeuralCMF), which is a self-supervised algorithm for modelling cardiac structures and delineating myocardial dynamics.

Traditional approaches, whether speckle-based or surface-based, face intrinsic limitations. The former faces challenges in tracking extended temporal frames, while the latter must contend with interpolative inaccuracies, especially evident in non-uniform myocardium movement cases. NeuralCMF overcomes these challenges, providing a holistic view of cardiac motion rather than a frame-by-frame analysis.
The burgeoning field of deep learning, while transformative, presents challenges, especially in data dependency. However, our NeuralCMF does not need extensive paired datasets. 
This is a crucial distinction from supervised approaches, as it allows our model to learn directly from the data, adhering to the inherent physical laws governing cardiac motion. The self-supervised nature of our approach depends on these universal physical principles, which all hearts follow, thus ensuring its generalizability. Every human heart exhibits similar fundamental physiological patterns and motions. By aligning our model with these principles, we enable it to predict and understand cardiac dynamics across different individuals. We have rigorously tested our method on 127 individuals, including both healthy subjects and patients with CHD. The fact that our method performs consistently across such a heterogeneous group demonstrates its potential for wide clinical application and its reliability across a broad spectrum of cardiac conditions. We provide additional results of myocardial motion tracking for LV, LA, and RA across diverse individuals in Fig. \ref{More_tracking_results3d}.

One of the remarkable features of NeuralCMF is its adaptability. The model accurately estimates myocardial motion when interpreting multi-view 2DE or 3DE videos or dealing with varied resolutions. Such adaptability, particularly its capacity to integrate with different imaging modalities, makes it a valuable tool for researchers and clinicians. 
The process of reconstructing 3D motion patterns from multiple 2D images is not only challenging but also immensely practical, especially important in settings like emergency departments, intensive care units, outpatient and preoperative clinics, as well as in medically underserved regions ranging from rural areas to low- and middle-income countries, extending to manned space missions. This methodology allows for the utilization of cost-effective handheld 2DE probes for capturing images from diverse angles. Subsequently, these images can be synthesized to reconstruct the 3D motion of the heart, thereby revealing its dynamic functions with high precision. We provide additional results of multi-view 2DE dataset in Fig. \ref{More_tracking_results3d} that underscore the efficacy of our motion estimation approach. These results illustrate that our methodology is capable of precisely reconstructing the 3D structure of the heart, alongside capturing the motion patterns.

In this study, we demonstrate the capability to classify CHD by employing 3D motion data extracted from 3D datasets. Looking ahead, we plan to build a model that utilize the 3D motion data extracted from multi-view 2DE to detect CHD. This model aims to be perfectly compatible with handheld 2D ultrasound devices, facilitating rapid and reliable screening for coronary heart disease. Such advancements hold the promise of revolutionizing the approach to cardiac care, especially in underserved regions, by making high-quality diagnostic procedures more accessible and efficient. Although our research primarily concentrates on cardiac imaging, the NeuralCMF framework—holds significant potential for broader applications. Specifically, this technique could be extended to the study of other organs and different imaging modalities, suggesting a wide range of future research directions.

The current computational time per subject presents a limitation for our proposed method, particularly as the study population increases. However, this issue can be effectively addressed through several strategies, categorized into hardware acceleration and algorithmic improvements. For hardware acceleration, we plan to adopt NVIDIA's tiny-cuda-nn framework instead of PyTorch can drastically reduce processing time, as demonstrated by the Instant-NGP \cite{muller2022instant}. Additionally, we intend to implement a distributed strategy across multiple GPUs, allowing each GPU to process a smaller portion of the data simultaneously, thereby decreasing the training time per iteration. On the algorithmic front, we aim to leverage meta-learning \cite{tancik2021learned} to learn an optimal initialization of model parameters from a large heart dataset. This approach can enhance both optimization speed and accuracy. Together, these proposed adjustments are expected to effectively reduce overall computational demands, thereby enhancing the feasibility of our method for practical applications in clinical and real-world settings.


\begin{thebibliography}{10}

\bibitem{nowbar2019mortality}
A.~N. Nowbar, M.~Gitto, J.~P. Howard, D.~P. Francis, and R.~Al-Lamee, ``Mortality from ischemic heart disease: Analysis of data from the world health organization and coronary artery disease risk factors from ncd risk factor collaboration,'' {\em Circulation: cardiovascular quality and outcomes}, vol.~12, no.~6, p.~e005375, 2019.

\bibitem{tsao2022heart}
C.~W. Tsao, A.~W. Aday, Z.~I. Almarzooq, A.~Alonso, A.~Z. Beaton, M.~S. Bittencourt, A.~K. Boehme, A.~E. Buxton, A.~P. Carson, Y.~Commodore-Mensah, {\em et~al.}, ``Heart disease and stroke statistics—2022 update: a report from the american heart association,'' {\em Circulation}, vol.~145, no.~8, pp.~e153--e639, 2022.

\bibitem{zhang2018abnormalities}
K.~W. Zhang, B.~S. Finkelman, G.~Gulati, H.~K. Narayan, J.~Upshaw, V.~Narayan, T.~Plappert, V.~Englefield, A.~M. Smith, C.~Zhang, {\em et~al.}, ``Abnormalities in 3-dimensional left ventricular mechanics with anthracycline chemotherapy are associated with systolic and diastolic dysfunction,'' {\em JACC: Cardiovascular Imaging}, vol.~11, no.~8, pp.~1059--1068, 2018.

\bibitem{van2019adding}
M.~J. van Mourik, D.~V. Zaar, M.~W. Smulders, J.~Heijman, J.~Lumens, J.~E. Dokter, V.~L. Passos, S.~Schalla, C.~Knackstedt, G.~Schummers, {\em et~al.}, ``Adding speckle-tracking echocardiography to visual assessment of systolic wall motion abnormalities improves the detection of myocardial infarction,'' {\em Journal of the American Society of Echocardiography}, vol.~32, no.~1, pp.~65--73, 2019.

\bibitem{alashi2018incremental}
A.~Alashi, A.~Mentias, A.~Abdallah, K.~Feng, A.~M. Gillinov, L.~L. Rodriguez, D.~R. Johnston, L.~G. Svensson, Z.~B. Popovic, B.~P. Griffin, {\em et~al.}, ``Incremental prognostic utility of left ventricular global longitudinal strain in asymptomatic patients with significant chronic aortic regurgitation and preserved left ventricular ejection fraction,'' {\em JACC: Cardiovascular Imaging}, vol.~11, no.~5, pp.~673--682, 2018.

\bibitem{kim2018myocardial}
H.~M. Kim, G.-Y. Cho, I.-C. Hwang, H.-M. Choi, J.-B. Park, Y.~E. Yoon, and H.-K. Kim, ``Myocardial strain in prediction of outcomes after surgery for severe mitral regurgitation,'' {\em JACC: Cardiovascular Imaging}, vol.~11, no.~9, pp.~1235--1244, 2018.

\bibitem{morris2017potential}
D.~Morris, E.~Belyavskiy, R.~Aravind-Kumar, M.~Kropf, A.~Frydas, K.~Braunauer, E.~Marquez, M.~Krisper, R.~Lindhorst, E.~Osmanoglou, {\em et~al.}, ``Potential usefulness and clinical relevance of adding left atrial strain to left atrial volume index in the detection of left ventricular diastolic dysfunction. jacc cardiovasc imaging. 2018; 11 (10): 1405--15,'' {\em J. JCMG}, vol.~29, 2017.

\bibitem{park2018global}
J.~J. Park, J.-B. Park, J.-H. Park, and G.-Y. Cho, ``Global longitudinal strain to predict mortality in patients with acute heart failure,'' {\em Journal of the American College of Cardiology}, vol.~71, no.~18, pp.~1947--1957, 2018.

\bibitem{risum2012simple}
N.~Risum, C.~Jons, N.~T. Olsen, T.~Fritz-Hansen, N.~E. Bruun, M.~V. Hojgaard, N.~Valeur, M.~B. Kronborg, J.~Kisslo, and P.~Sogaard, ``Simple regional strain pattern analysis to predict response to cardiac resynchronization therapy: rationale, initial results, and advantages,'' {\em American Heart Journal}, vol.~163, no.~4, pp.~697--704, 2012.

\bibitem{fine2013outcome}
N.~M. Fine, L.~Chen, P.~M. Bastiansen, R.~P. Frantz, P.~A. Pellikka, J.~K. Oh, and G.~C. Kane, ``Outcome prediction by quantitative right ventricular function assessment in 575 subjects evaluated for pulmonary hypertension,'' {\em Circulation: Cardiovascular Imaging}, vol.~6, no.~5, pp.~711--721, 2013.

\bibitem{chowdhury2015speckle}
S.~M. Chowdhury, Z.~M. Hijazi, J.~T. Fahey, J.~F. Rhodes, S.~Kar, R.~Makkar, M.~Mullen, Q.-L. Cao, and G.~S. Shirali, ``Speckle-tracking echocardiographic measures of right ventricular function correlate with improvement in exercise function after percutaneous pulmonary valve implantation,'' {\em Journal of the American Society of Echocardiography}, vol.~28, no.~9, pp.~1036--1044, 2015.

\bibitem{leung2018left}
M.~Leung, P.~J. van Rosendael, R.~Abou, N.~Ajmone~Marsan, D.~Y. Leung, V.~Delgado, and J.~J. Bax, ``Left atrial function to identify patients with atrial fibrillation at high risk of stroke: new insights from a large registry,'' {\em European heart journal}, vol.~39, no.~16, pp.~1416--1425, 2018.

\bibitem{motoki2014global}
H.~Motoki, K.~Negishi, K.~Kusunose, Z.~B. Popovi{\'c}, M.~Bhargava, O.~M. Wazni, W.~I. Saliba, M.~K. Chung, T.~H. Marwick, and A.~L. Klein, ``Global left atrial strain in the prediction of sinus rhythm maintenance after catheter ablation for atrial fibrillation,'' {\em Journal of the American Society of Echocardiography}, vol.~27, no.~11, pp.~1184--1192, 2014.

\bibitem{braga2019trends}
J.~R. Braga, H.~Leong-Poi, V.~E. Rac, P.~C. Austin, H.~J. Ross, and D.~S. Lee, ``Trends in the use of cardiac imaging for patients with heart failure in canada,'' {\em JAMA network open}, vol.~2, no.~8, pp.~e198766--e198766, 2019.

\bibitem{liu2019deep}
S.~Liu, Y.~Wang, X.~Yang, B.~Lei, L.~Liu, S.~X. Li, D.~Ni, and T.~Wang, ``Deep learning in medical ultrasound analysis: a review,'' {\em Engineering}, vol.~5, no.~2, pp.~261--275, 2019.

\bibitem{voigt20192}
J.-U. Voigt and M.~Cvijic, ``2-and 3-dimensional myocardial strain in cardiac health and disease,'' {\em JACC: Cardiovascular Imaging}, vol.~12, no.~9, pp.~1849--1863, 2019.

\bibitem{mondillo2011speckle}
S.~Mondillo, M.~Galderisi, D.~Mele, M.~Cameli, V.~S. Lomoriello, V.~Zac{\`a}, P.~Ballo, A.~D'Andrea, D.~Muraru, M.~Losi, {\em et~al.}, ``Speckle-tracking echocardiography: a new technique for assessing myocardial function,'' {\em Journal of Ultrasound in Medicine}, vol.~30, no.~1, pp.~71--83, 2011.

\bibitem{lubinski1999speckle}
M.~A. Lubinski, S.~Y. Emelianov, and M.~O'Donnell, ``Speckle tracking methods for ultrasonic elasticity imaging using short-time correlation,'' {\em IEEE transactions on ultrasonics, ferroelectrics, and frequency control}, vol.~46, no.~1, pp.~82--96, 1999.

\bibitem{chen20053}
X.~Chen, H.~Xie, R.~Erkamp, K.~Kim, C.~Jia, J.~Rubin, and M.~O'Donnell, ``3-d correlation-based speckle tracking,'' {\em Ultrasonic Imaging}, vol.~27, no.~1, pp.~21--36, 2005.

\bibitem{papademetris2002estimation}
X.~Papademetris, A.~J. Sinusas, D.~P. Dione, R.~T. Constable, and J.~S. Duncan, ``Estimation of 3-d left ventricular deformation from medical images using biomechanical models,'' {\em IEEE transactions on medical imaging}, vol.~21, no.~7, pp.~786--800, 2002.

\bibitem{parajuli2019flow}
N.~Parajuli, A.~Lu, K.~Ta, J.~Stendahl, N.~Boutagy, I.~Alkhalil, M.~Eberle, G.-S. Jeng, M.~Zontak, M.~O’Donnell, {\em et~al.}, ``Flow network tracking for spatiotemporal and periodic point matching: Applied to cardiac motion analysis,'' {\em Medical image analysis}, vol.~55, pp.~116--135, 2019.

\bibitem{huang2014contour}
X.~Huang, D.~P. Dione, C.~B. Compas, X.~Papademetris, B.~A. Lin, A.~Bregasi, A.~J. Sinusas, L.~H. Staib, and J.~S. Duncan, ``Contour tracking in echocardiographic sequences via sparse representation and dictionary learning,'' {\em Medical image analysis}, vol.~18, no.~2, pp.~253--271, 2014.

\bibitem{parajuli2016integrated}
N.~Parajuli, A.~Lu, J.~C. Stendahl, M.~Zontak, N.~Boutagy, M.~Eberle, I.~Alkhalil, M.~O’Donnell, A.~J. Sinusas, and J.~S. Duncan, ``Integrated dynamic shape tracking and rf speckle tracking for cardiac motion analysis,'' in {\em MICCAI}, pp.~431--438, Springer, 2016.

\bibitem{shi2000point}
P.~Shi, A.~J. Sinusas, R.~T. Constable, E.~Ritman, and J.~S. Duncan, ``Point-tracked quantitative analysis of left ventricular surface motion from 3-d image sequences,'' {\em IEEE transactions on medical imaging}, vol.~19, no.~1, pp.~36--50, 2000.

\bibitem{lin2004generalized}
N.~Lin and J.~S. Duncan, ``Generalized robust point matching using an extended free-form deformation model: application to cardiac images,'' in {\em 2004 2nd IEEE International Symposium on Biomedical Imaging: Nano to Macro (IEEE Cat No. 04EX821)}, pp.~320--323, IEEE, 2004.

\bibitem{blankstein2016evaluation}
R.~Blankstein and A.~H. Waller, ``Evaluation of known or suspected cardiac sarcoidosis,'' {\em Circulation: Cardiovascular Imaging}, vol.~9, no.~3, p.~e000867, 2016.

\bibitem{ichinose2008mri}
A.~Ichinose, H.~Otani, M.~Oikawa, K.~Takase, H.~Saito, H.~Shimokawa, and S.~Takahashi, ``Mri of cardiac sarcoidosis: basal and subepicardial localization of myocardial lesions and their effect on left ventricular function,'' {\em American Journal of Roentgenology}, vol.~191, no.~3, pp.~862--869, 2008.

\bibitem{ahn2020unsupervised}
S.~S. Ahn, K.~Ta, A.~Lu, J.~C. Stendahl, A.~J. Sinusas, and J.~S. Duncan, ``Unsupervised motion tracking of left ventricle in echocardiography,'' in {\em Medical imaging 2020: Ultrasonic imaging and tomography}, vol.~11319, pp.~196--202, SPIE, 2020.

\bibitem{ta2020semi}
K.~Ta, S.~S. Ahn, J.~C. Stendahl, A.~J. Sinusas, and J.~S. Duncan, ``A semi-supervised joint network for simultaneous left ventricular motion tracking and segmentation in 4d echocardiography,'' in {\em MICCAI}, pp.~468--477, Springer, 2020.

\bibitem{ahn2023co}
S.~S. Ahn, K.~Ta, S.~L. Thorn, J.~A. Onofrey, I.~H. Melvinsdottir, S.~Lee, J.~Langdon, A.~J. Sinusas, and J.~S. Duncan, ``Co-attention spatial transformer network for unsupervised motion tracking and cardiac strain analysis in 3d echocardiography,'' {\em Medical Image Analysis}, vol.~84, p.~102711, 2023.

\bibitem{balakrishnan2019voxelmorph}
G.~Balakrishnan, A.~Zhao, M.~R. Sabuncu, J.~Guttag, and A.~V. Dalca, ``Voxelmorph: a learning framework for deformable medical image registration,'' {\em IEEE transactions on medical imaging}, vol.~38, no.~8, pp.~1788--1800, 2019.

\bibitem{alessandrini2015pipeline}
M.~Alessandrini, M.~De~Craene, O.~Bernard, S.~Giffard-Roisin, P.~Allain, I.~Waechter-Stehle, J.~Weese, E.~Saloux, H.~Delingette, M.~Sermesant, {\em et~al.}, ``A pipeline for the generation of realistic 3d synthetic echocardiographic sequences: Methodology and open-access database,'' {\em IEEE transactions on medical imaging}, vol.~34, no.~7, pp.~1436--1451, 2015.

\bibitem{besl1992method}
P.~J. Besl and N.~D. McKay, ``Method for registration of 3-d shapes,'' in {\em Sensor fusion IV: control paradigms and data structures}, vol.~1611, pp.~586--606, Spie, 1992.

\bibitem{chui2003new}
H.~Chui and A.~Rangarajan, ``A new point matching algorithm for non-rigid registration,'' {\em Computer Vision and Image Understanding}, vol.~89, no.~2-3, pp.~114--141, 2003.

\bibitem{rohe2017svf}
M.-M. Roh{\'e}, M.~Datar, T.~Heimann, M.~Sermesant, and X.~Pennec, ``Svf-net: learning deformable image registration using shape matching,'' in {\em MICCAI}, pp.~266--274, Springer, 2017.

\bibitem{wu2018deep}
J.~Wu, T.~R. Mazur, S.~Ruan, C.~Lian, N.~Daniel, H.~Lashmett, L.~Ochoa, I.~Zoberi, M.~A. Anastasio, H.~M. Gach, {\em et~al.}, ``A deep boltzmann machine-driven level set method for heart motion tracking using cine mri images,'' {\em Medical image analysis}, vol.~47, pp.~68--80, 2018.

\bibitem{ronneberger2015u}
O.~Ronneberger, P.~Fischer, and T.~Brox, ``U-net: Convolutional networks for biomedical image segmentation,'' in {\em MICCAI}, pp.~234--241, Springer, 2015.

\bibitem{yu2020foal}
H.~Yu, S.~Sun, H.~Yu, X.~Chen, H.~Shi, T.~S. Huang, and T.~Chen, ``Foal: Fast online adaptive learning for cardiac motion estimation,'' in {\em IEEE CVPR}, pp.~4313--4323, 2020.

\bibitem{yu2020motion}
H.~Yu, X.~Chen, H.~Shi, T.~Chen, T.~S. Huang, and S.~Sun, ``Motion pyramid networks for accurate and efficient cardiac motion estimation,'' in {\em MICCAI}, pp.~436--446, Springer, 2020.

\bibitem{sitzmann2020implicit}
V.~Sitzmann, J.~Martel, A.~Bergman, D.~Lindell, and G.~Wetzstein, ``Implicit neural representations with periodic activation functions,'' {\em NeurIPS}, vol.~33, pp.~7462--7473, 2020.

\bibitem{mildenhall2021nerf}
B.~Mildenhall, P.~P. Srinivasan, M.~Tancik, J.~T. Barron, R.~Ramamoorthi, and R.~Ng, ``Nerf: Representing scenes as neural radiance fields for view synthesis,'' {\em Communications of the ACM}, vol.~65, no.~1, pp.~99--106, 2021.

\bibitem{he2016deep}
K.~He, X.~Zhang, S.~Ren, and J.~Sun, ``Deep residual learning for image recognition,'' in {\em IEEE CVPR}, pp.~770--778, 2016.

\bibitem{zhu2023pyramid}
J.~Zhu, H.~Zhu, Q.~Zhang, F.~Zhu, Z.~Ma, and X.~Cao, ``Pyramid nerf: Frequency guided fast radiance field optimization,'' {\em International Journal of Computer Vision}, pp.~1--16, 2023.

\bibitem{liu2022recovery}
R.~Liu, Y.~Sun, J.~Zhu, L.~Tian, and U.~S. Kamilov, ``Recovery of continuous 3d refractive index maps from discrete intensity-only measurements using neural fields,'' {\em Nature Machine Intelligence}, vol.~4, no.~9, pp.~781--791, 2022.

\bibitem{zhu2022dnf}
H.~Zhu, Z.~Liu, Y.~Zhou, Z.~Ma, and X.~Cao, ``Dnf: diffractive neural field for lensless microscopic imaging,'' {\em Optics Express}, vol.~30, no.~11, pp.~18168--18178, 2022.

\bibitem{chen2020physics}
Y.~Chen, L.~Lu, G.~E. Karniadakis, and L.~Dal~Negro, ``Physics-informed neural networks for inverse problems in nano-optics and metamaterials,'' {\em Optics express}, vol.~28, no.~8, pp.~11618--11633, 2020.

\bibitem{karniadakis2021physics}
G.~E. Karniadakis, I.~G. Kevrekidis, L.~Lu, P.~Perdikaris, S.~Wang, and L.~Yang, ``Physics-informed machine learning,'' {\em Nature Reviews Physics}, vol.~3, no.~6, pp.~422--440, 2021.

\bibitem{shen2022nerp}
L.~Shen, J.~Pauly, and L.~Xing, ``Nerp: implicit neural representation learning with prior embedding for sparsely sampled image reconstruction,'' {\em IEEE Transactions on Neural Networks and Learning Systems}, 2022.

\bibitem{zha2022naf}
R.~Zha, Y.~Zhang, and H.~Li, ``Naf: neural attenuation fields for sparse-view cbct reconstruction,'' in {\em MICCAI}, pp.~442--452, Springer, 2022.

\bibitem{wu2022arbitrary}
Q.~Wu, Y.~Li, Y.~Sun, Y.~Zhou, H.~Wei, J.~Yu, and Y.~Zhang, ``An arbitrary scale super-resolution approach for 3d mr images via implicit neural representation,'' {\em IEEE Journal of Biomedical and Health Informatics}, vol.~27, no.~2, pp.~1004--1015, 2022.

\bibitem{shen2023cardiacfield}
C.~Shen, H.~Zhu, Y.~Zhou, Y.~Liu, S.~Yi, L.~Dong, W.~Zhao, D.~Brady, X.~Cao, Z.~Ma, and Y.~Lin, ``Cardiacfield: Computational echocardiography for universal screening,'' {\em Research Square}, 2023.

\bibitem{muller2022instant}
T.~M{\"u}ller, A.~Evans, C.~Schied, and A.~Keller, ``Instant neural graphics primitives with a multiresolution hash encoding,'' {\em ACM Transactions on Graphics (ToG)}, vol.~41, no.~4, pp.~1--15, 2022.

\bibitem{zhu2023disorder}
H.~Zhu, S.~Xie, Z.~Liu, F.~Liu, Q.~Zhang, Y.~Zhou, Y.~Lin, Z.~Ma, and X.~Cao, ``Disorder-invariant implicit neural representation,'' {\em IEEE Transactions on Pattern Analysis and Machine Intelligence}, 2024.

\bibitem{de2012temporal}
M.~De~Craene, G.~Piella, O.~Camara, N.~Duchateau, E.~Silva, A.~Doltra, J.~D’hooge, J.~Brugada, M.~Sitges, and A.~F. Frangi, ``Temporal diffeomorphic free-form deformation: Application to motion and strain estimation from 3d echocardiography,'' {\em Medical image analysis}, vol.~16, no.~2, pp.~427--450, 2012.

\bibitem{kingma2014adam}
D.~P. Kingma and J.~Ba, ``Adam: A method for stochastic optimization,'' {\em arXiv preprint arXiv:1412.6980}, 2014.

\bibitem{ostenfeld2012manual}
E.~Ostenfeld, M.~Carlsson, K.~Shahgaldi, A.~Roijer, and J.~Holm, ``Manual correction of semi-automatic three-dimensional echocardiography is needed for right ventricular assessment in adults; validation with cardiac magnetic resonance,'' {\em Cardiovascular Ultrasound}, vol.~10, pp.~1--10, 2012.

\bibitem{seo2020right}
Y.~Seo, T.~Ishizu, M.~Ieda, and N.~Ohte, ``Right ventricular three-dimensional echocardiography: the current status and future perspectives,'' {\em Journal of Echocardiography}, vol.~18, pp.~149--159, 2020.

\bibitem{krebs2019learning}
J.~Krebs, H.~Delingette, B.~Mailh{\'e}, N.~Ayache, and T.~Mansi, ``Learning a probabilistic model for diffeomorphic registration,'' {\em IEEE transactions on medical imaging}, vol.~38, no.~9, pp.~2165--2176, 2019.

\bibitem{tancik2021learned}
M.~Tancik, B.~Mildenhall, T.~Wang, D.~Schmidt, P.~P. Srinivasan, J.~T. Barron, and R.~Ng, ``Learned initializations for optimizing coordinate-based neural representations,'' in {\em Proceedings of the IEEE/CVF Conference on Computer Vision and Pattern Recognition}, pp.~2846--2855, 2021.

\end{thebibliography}

\end{document}